\documentclass[numbib]{imaiai-plain}

\PassOptionsToPackage{natbib}{numbers}

\jno{XXXXXXX}
\received{\today}

\newcommand*{\tikzextpath}{tikz/}

\usepackage[utf8]{inputenc}
\usepackage{import}

\ifdefined\uselmodern
\usepackage{lmodern}
\fi

\usepackage[T1]{fontenc}
\usepackage{etex}
\usepackage{color}
\usepackage{microtype}
\usepackage{etoolbox}\usepackage{ifthen}
\usepackage{xparse}
\usepackage{substr}
\usepackage{subcaption}
\usepackage[hide=true,dir=bashful]{bashful}
\usepackage{changepage}
\usepackage{xspace}

\makeatletter
\@ifclassloaded{imaiai}{}{
  \@ifclassloaded{imaiai-plain}{}{
    \@ifclassloaded{IEEEtran}{
          }{
      \@ifundefined{nobib}{
        \usepackage{biblatex}
        \addbibresource{IEEEabrv.bib}
        \addbibresource{literature.bib}
      }
    }
  }
}
\makeatother

\usepackage{mathtools}

\usepackage{booktabs}
\usepackage{array}

\usepackage{dsfont}
\usepackage{mathrsfs}

\usepackage{setspace}

\usepackage[justification=centering]{caption}

\usepackage{colonequals}
\usepackage{centernot}
\usepackage{relsize}
\usepackage{tikz}
\usepackage{pgfplots}
\usepackage{longtable}
\usepackage{eqparbox}
\usepackage{fp}

\usepackage{comment}

\makeatletter
\@ifclassloaded{beamer}{
  \usepackage{appendixnumberbeamer}
}{
  \usepackage{amssymb}
  \usepackage{environ}
  \usepackage{amsmath}
  \usepackage[bottom]{footmisc}
  \usepackage[thepage]{zref}
  \usepackage{enumitem}
  \@ifclassloaded{IEEEtran}{}{
    \@ifclassloaded{a0poster}{}{
      \@ifpackageloaded{classicthesis}{}{
        \@ifclassloaded{imaiai}{}{
          \@ifclassloaded{imaiai-plain}{}{
            \usepackage[heightrounded=true]{geometry}
          }
        }
      }
    }
    \usepackage[nottoc]{tocbibind}
    \usepackage{imakeidx}
  }
  \@ifundefined{nolinks}{
    \usepackage{hyperref}
    \@ifclassloaded{a0poster}{}{
      \usepackage[all]{hypcap}
    }
    \usepackage{bookmark}
  }{
    \newcommand{\texorpdfstring}[2]{#1}
  }
}
\makeatother

\usepackage{amsthm}
\usepackage{thmtools, thm-restate}

\ifdefined\crefnocapitalise
\usepackage{cleveref}
\else
\usepackage[capitalise]{cleveref}
\fi

\ifdefined\useautonum
\makeatletter
\@ifclassloaded{beamer}{
  \usepackage{scrpage2}
}{
    \usepackage{autonum}
  \ifdef{\Cref}{%
    \autonum@generatePatchedReferenceCSL{Cref}%
  }{}%
}
\makeatother
\fi

\usepackage{wasysym}

\usepackage{fixmath}

\crefname{apx}{appendix}{appendices}
\Crefname{apx}{Appendix}{Appendices}
\crefformat{apx}{the #2appendix#3}
\Crefformat{apx}{the #2Appendix#3}

\newtheorem*{conjecture*}{Conjecture}
\newtheorem*{definition*}{Definition}
\newtheorem*{proposition*}{Proposition}
\newtheorem*{corollary*}{Corollary}
\makeatletter
\@ifclassloaded{beamer}{
  \newtheorem{conjecture}{Conjecture}
  \newtheorem*{lemma*}{Lemma}
  \newtheorem*{theorem*}{Theorem}
  \newtheorem*{example*}{Example}
}{
  \newcommand{\mynewtheorem}[2]{
    \ifdef{\seperatenumbering}{
      \newtheorem{#1}{#2}
    }{
      \newtheorem{#1}[theorem]{#2}
    }
  }
  \@ifclassloaded{imaiai}{}{
    \@ifclassloaded{imaiai-plain}{}{
      \ifdef{\numbertheoremwithin}{
        \newtheorem{theorem}{Theorem}[\numbertheoremwithin]
      }{
        \newtheorem{theorem}{Theorem}
      }

      \newtheorem*{theorem*}{Theorem}
      
      \mynewtheorem{lemma}{Lemma}
      \mynewtheorem{corollary}{Corollary}
      \mynewtheorem{definition}{Definition}
      \mynewtheorem{proposition}{Proposition}
      \mynewtheorem{conjecture}{Conjecture}
      
      \theoremstyle{remark}
      
      \newtheorem{remark}{Remark}
      \newtheorem*{remark*}{Remark}
      \newtheorem*{example*}{Example}
    }
  }
}
\makeatother

\crefname{conjecture}{Conjecture}{Conjectures}
\Crefname{conjecture}{Conjecture}{Conjectures}

\crefname{enumi}{part}{parts}
\Crefname{enumi}{Part}{Parts}

\crefname{equation}{Equation}{Equations}
\Crefname{equation}{Equation}{Equations}
\crefformat{equation}{#2(#1)#3}
\crefrangeformat{equation}{#3(#1)#4--#5(#2)#6}
\crefmultiformat{equation}{#2(#1)#3}{ and~#2(#1)#3}{, #2(#1)#3}{ and~#2(#1)#3}
\Crefformat{equation}{#2(#1)#3}
\Crefrangeformat{equation}{#3(#1)#4--#5(#2)#6}
\Crefmultiformat{equation}{#2(#1)#3}{ and~#2(#1)#3}{, #2(#1)#3}{ and~#2(#1)#3}

\crefname{page}{page}{page}
\Crefname{page}{Page}{Page}

\crefname{assumption}{assumption}{assumptions}
\Crefname{assumption}{Assumption}{Assumptions}

\crefname{def}{definition}{definition}
\Crefname{def}{Definition}{Definition}

\crefname{case}{case}{cases}
\Crefname{case}{Case}{Cases}

\let\originalleft\left
\let\originalright\right
\let\originalmiddle\middle
\renewcommand{\left}{\mathopen{}\mathclose\bgroup\originalleft}
\renewcommand{\right}{\aftergroup\egroup\originalright}
\renewcommand{\middle}{\originalmiddle}

\makeatletter
\@ifclassloaded{beamer}{%
  \newcommand{\todo}[1][\empty]{%
    \textbf{\textcolor{red}{\ifthenelse{\equal{#1}{\empty}}{TODO}{TODO:~\emph{#1}}}}%
  }%
}
{
  \newcounter{todo}
  \newcommand{\todo}[1][\empty]{%
    \bookmarksetupnext{keeplevel}    \subpdfbookmark{TODO}{todo.\arabic{todo}}%
    \bookmarksetup{keeplevel=false}%
    \refstepcounter{todo}%
    \textcolor{red}{\textbf{TODO}%
      \ifthenelse{\equal{#1}{\empty}}{}%
      {\footnote{\textcolor{red}{\emph{#1}}}}%
    }%
    \xspace%
  }
}
\makeatother

\makeatletter
\@ifclassloaded{beamer}{
  \newcommand\changemode[1]{%
    \gdef\beamer@currentmode{#1}}
}{}
\makeatother

\makeatletter
\@ifclassloaded{IEEEtran}{
  
}{}
\makeatother

\newcommand{\stacklap}[2]{\stackrel{\mathclap{#1}}{#2}}

\newcounter{align}
\expandafter\let\expandafter\oldalign\csname align\endcsname
\expandafter\def\csname align\endcsname{\refstepcounter{align}\oldalign}
\newcommand{\RA}[2][\empty]{\stackrel{\eqmakebox[\thealign]{\text{\scriptsize#1}}}{#2}}
\newcommand{\RL}[2][\empty]{\stackrel{\mathclap{\text{\scriptsize#1}}}{#2}}

\newcommand{\ie}{i.e.\@\xspace}

\newcommand{\eg}{e.g.\@\xspace}
\newcommand{\Eg}{For example\@\xspace}
\newcommand{\cf}{cf.\@\xspace}

\newcommand{\etal}{\emph{et~al.}\@\xspace}

\newcommand{\iid}{i.i.d.\@\xspace}

\newcommand{\eqnl}{\nonumber\\*}

\newcommand{\prm}{^\prime}
\newcommand{\pprm}{^{\prime\prime}}

\DeclareDocumentCommand{\card}{ O{\empty} m}{
  \ensuremath{
    \ifthenelse{\equal{#1}{big}}{\big\vert #2 \big\vert}{
      \ifthenelse{\equal{#1}{Big}}{\Big\vert #2 \Big\vert}{
        \ifthenelse{\equal{#1}{bigg}}{\bigg\vert #2 \bigg\vert}{
          \ifthenelse{\equal{#1}{Bigg}}{\Bigg\vert #2 \Bigg\vert}{
            \ifthenelse{\equal{#1}{normal}}{\vert #2 \vert}{\left\vert #2 \right\vert}
          }
        }
      }
    }
  }
}

\DeclareDocumentCommand{\Vol}{ O{\empty} m}{
  \ensuremath{
    \ensuremath{\mathrm{Vol}
      \ifthenelse{\equal{#1}{big}}{\big(#2\big)}{
        \ifthenelse{\equal{#1}{Big}}{\Big(#2\Big)}{
          \ifthenelse{\equal{#1}{bigg}}{\bigg(#2\bigg)}{
            \ifthenelse{\equal{#1}{Bigg}}{\Bigg(#2\Bigg)}{
              \ifthenelse{\equal{#1}{normal}}{(#2)}{\left(#2\right)}
            }
          }
        }
      }
    }
  }
}

\newcommand{\cc}[2][\empty]{\ensuremath{\mathrm{cc}_{#1}(#2)}}
\newcommand{\ext}[1]{\ensuremath{\mathrm{ext}(#1)}}
\DeclareDocumentCommand{\cvx}{ O{\empty} m}{
  \ensuremath{
    \ensuremath{\mathrm{conv}
      \ifthenelse{\equal{#1}{big}}{\big(#2\big)}{
        \ifthenelse{\equal{#1}{Big}}{\Big(#2\Big)}{
          \ifthenelse{\equal{#1}{bigg}}{\bigg(#2\bigg)}{
            \ifthenelse{\equal{#1}{Bigg}}{\Bigg(#2\Bigg)}{
              \ifthenelse{\equal{#1}{normal}}{(#2)}{\left(#2\right)}
            }
          }
        }
      }
    }
  }
}

\newcommand{\transp}{\ensuremath{^{\mathrm{T}}}}

\newcommand{\ol}[1]{\ensuremath{\overline{#1}}}

\DeclareDocumentCommand{\Ntoo}{ O{\empty} m}{
  \ensuremath{
    \ifthenelse{\equal{#1}{big}}{\big[#2\big]}{
      \ifthenelse{\equal{#1}{Big}}{\Big[#2\Big]}{
        \ifthenelse{\equal{#1}{bigg}}{\bigg[#2\bigg]}{
          \ifthenelse{\equal{#1}{Bigg}}{\Bigg[#2\Bigg]}{
            \ifthenelse{\equal{#1}{normal}}{[#2]}{\left[#2\right]}
          }
        }
      }
    }
  }
}
\DeclareDocumentCommand{\NZtoo}{ O{\empty} m}{
  \ensuremath{
    \ifthenelse{\equal{#1}{big}}{\big[0\,{:}\,#2\big]}{
      \ifthenelse{\equal{#1}{Big}}{\Big[0\,{:}\,#2\Big]}{
        \ifthenelse{\equal{#1}{bigg}}{\bigg[0\,{:}\,#2\bigg]}{
          \ifthenelse{\equal{#1}{Bigg}}{\Bigg[0\,{:}\,#2\Bigg]}{
            \ifthenelse{\equal{#1}{normal}}{[0\,{:}\,#2]}{\left[0\,{:}\,#2\right]}
          }
        }
      }
    }
  }
}

\newcommand{\defas}{\vcentcolon=}

\newcommand{\ee}[1]{\ensuremath{\mathrm{e}^{#1}}}

\newcommand{\wt}[1]{\ensuremath{\widetilde{#1}}}

\newcommand{\NN}{\ensuremath{\mathbb{N}}}

\newcommand{\RR}{\ensuremath{\mathbb{R}}}

\newcommand{\vt}[1]{\ensuremath{\boldsymbol{#1}}}

\newcommand{\wc}{\ensuremath{{}\cdot{}}}

\DeclareDocumentCommand{\ind}{m m}{%
  \ensuremath{%
    \mathds{1}_{#1}%
    \ifthenelse{\equal{#2}{\empty}}{}{(#2)}%
  }%
}

\DeclareDocumentCommand{\Exp}{ O{\empty} O{\empty} m}{
  \ensuremath{\mathds{E}_{#1}
    \ifthenelse{\equal{#2}{big}}{\big[ #3 \big]}{
      \ifthenelse{\equal{#2}{Big}}{\Big[ #3 \Big]}{
        \ifthenelse{\equal{#2}{bigg}}{\bigg[ #3 \bigg]}{
          \ifthenelse{\equal{#2}{Bigg}}{\Bigg[ #3 \Bigg]}{
            \ifthenelse{\equal{#2}{normal}}{[ #3 ]}{\left[ #3 \right]}
          }
        }
      }
    }
  }
}

\DeclareDocumentCommand{\condExp}{ O{\empty} m m}{
  \ensuremath{\mathds{E}
    \ifthenelse{\equal{#1}{big}}{\big[{#2\big| #3}\big]}{
      \ifthenelse{\equal{#1}{Big}}{\Big[{#2\Big| #3}\Big]}{
        \ifthenelse{\equal{#1}{bigg}}{\bigg[{#2\bigg| #3}\bigg]}{
          \ifthenelse{\equal{#1}{Bigg}}{\Bigg[{#2\Bigg| #3}\Bigg]}{
            \ifthenelse{\equal{#1}{normal}}{[{#2| #3}]}{\left[#2 \middle| #3\right]}
          }
        }
      }
    }
  }
}

\newcommand{\p}[2][\empty]{\ensuremath{\mathrm{p}_{#1}\ifthenelse{\equal{#2}{}}{}{\left({#2}\right)}}}
\newcommand{\q}[2][\empty]{\ensuremath{\mathrm{q}_{#1}\ifthenelse{\equal{#2}{}}{}{\left({#2}\right)}}}
\newcommand{\wtp}[2][\empty]{\ensuremath{\wt{\mathrm{p}}_{#1}\ifthenelse{\equal{#2}{}}{}{\left({#2}\right)}}}
\newcommand{\htp}[2][\empty]{\ensuremath{\hat{\mathrm{p}}_{#1}\ifthenelse{\equal{#2}{}}{}{\left({#2}\right)}}}
\newcommand{\pp}[2][\empty]{\ensuremath{\mathrm{p}'_{#1}\ifthenelse{\equal{#2}{}}{}{\left({#2}\right)}}}
\newcommand{\qq}[2][\empty]{\ensuremath{\mathrm{q}'_{#1}\ifthenelse{\equal{#2}{}}{}{\left({#2}\right)}}}
\newcommand{\ppp}[2][\empty]{\ensuremath{\mathrm{p}''_{#1}\ifthenelse{\equal{#2}{}}{}{\left({#2}\right)}}}

\newcommand{\pdf}[2][\empty]{\ensuremath{\mathrm{f}_{#1}\ifthenelse{\equal{#2}{}}{}{\left({#2}\right)}}}

\DeclareDocumentCommand{\Prob}{ O{\empty} O{\empty} m}{
  \ensuremath{\mathrm{P}\ifthenelse{\equal{#1}{\empty}}{}{_{#1}}
    \ifthenelse{\equal{#2}{big}}{\big\{#3\big\}}{
      \ifthenelse{\equal{#2}{Big}}{\Big\{#3\Big\}}{
        \ifthenelse{\equal{#2}{bigg}}{\bigg\{#3\bigg\}}{
          \ifthenelse{\equal{#2}{Bigg}}{\Bigg\{#3\Bigg\}}{
            \ifthenelse{\equal{#2}{normal}}{\{#3\}}{\left\{#3\right\}}
          }
        }
      }
    }
  }
}

\DeclareDocumentCommand{\DKL}{ O{\empty} m m}{
  \ensuremath{\mathrm{D}
    \ifthenelse{\equal{#1}{big}}{\big(#2 \big\Vert #3\big)}{
      \ifthenelse{\equal{#1}{Big}}{\Big(#2 \Big\Vert #3\Big)}{
        \ifthenelse{\equal{#1}{bigg}}{\bigg(#2 \bigg\Vert #3\bigg)}{
          \ifthenelse{\equal{#1}{Bigg}}{\Bigg(#2 \Bigg\Vert #3\Bigg)}{
            \ifthenelse{\equal{#1}{normal}}{(#2 \Vert #3)}{\left(#2 \middle\Vert #3\right)}
          }
        }
      }
    }
  }
}

\DeclareDocumentCommand{\pcond}{ O{\empty} O{\empty} O{\empty} m m}{
  \ensuremath{\mathrm{p}
    \ifthenelse{\equal{#1#2}{\empty}}{}{_{#1|#2}}
    \ifthenelse{\equal{#4#5}{\empty}}{}{
      \ifthenelse{\equal{#3}{big}}{\big(#4 \big| #5\big)}{
        \ifthenelse{\equal{#3}{Big}}{\Big(#4 \Big| #5\Big)}{
          \ifthenelse{\equal{#3}{bigg}}{\bigg(#4 \bigg| #5\bigg)}{
            \ifthenelse{\equal{#3}{Bigg}}{\Bigg(#4 \Bigg| #5\Bigg)}{
              \ifthenelse{\equal{#3}{normal}}{(#4 | #5)}{\left(#4 \middle| #5\right)}
            }
          }
        }
      }
    }
  }
}

\DeclareDocumentCommand{\Pcond}{ O{\empty} O{\empty} O{\empty} m m}{
  \ensuremath{\mathrm{P}
    \ifthenelse{\equal{#1#2}{\empty}}{}{_{#1|#2}}
    \ifthenelse{\equal{#4#5}{\empty}}{}{
      \ifthenelse{\equal{#3}{big}}{\big\{#4 \big| #5\big\}}{
        \ifthenelse{\equal{#3}{Big}}{\Big\{#4 \Big| #5\Big\}}{
          \ifthenelse{\equal{#3}{bigg}}{\bigg\{#4 \bigg| #5\bigg\}}{
            \ifthenelse{\equal{#3}{Bigg}}{\Bigg\{#4 \Bigg| #5\Bigg\}}{
              \ifthenelse{\equal{#3}{normal}}{\{#4 | #5\}}{\left\{#4 \middle| #5\right\}}
            }
          }
        }
      }
    }
  }
}

\newcommand{\rv}[1]{\ensuremath{\uppercase{#1}}}
\newcommand{\rvt}[1]{\rv{\vt{#1}}} 
\newcommand{\wrv}[1]{\wt{\rv{#1}}}  
\newcommand{\hrv}[1]{\hat{\rv{#1}}}  

\DeclareDocumentCommand{\mutInf}{ O{\empty} m m}{
  \ensuremath{\mathrm{I}
    \ifthenelse{\equal{#1}{big}}{\big({#2; #3}\big)}{
      \ifthenelse{\equal{#1}{Big}}{\Big({#2; #3}\Big)}{
        \ifthenelse{\equal{#1}{bigg}}{\bigg({#2; #3}\bigg)}{
          \ifthenelse{\equal{#1}{Bigg}}{\Bigg({#2; #3}\Bigg)}{
            \ifthenelse{\equal{#1}{normal}}{({#2; #3})}{\left({#2; #3}\right)}
          }
        }
      }
    }
  }
}

\DeclareDocumentCommand{\condMutInf}{ O{\empty} m m m }{
  \ensuremath{\mathrm{I}
    \ifthenelse{\equal{#1}{big}}{\big(#2; #3 \big| #4\big)}{
      \ifthenelse{\equal{#1}{Big}}{\Big(#2; #3 \Big| #4\Big)}{
        \ifthenelse{\equal{#1}{bigg}}{\bigg(#2; #3 \bigg| #4\bigg)}{
          \ifthenelse{\equal{#1}{Bigg}}{\Bigg(#2; #3 \Bigg| #4\Bigg)}{
            \ifthenelse{\equal{#1}{normal}}{(#2; #3 | #4)}{\left(#2; #3 \middle| #4\right)}
          }
        }
      }
    }
  }
}

\DeclareDocumentCommand{\wtyp}{ O{\empty} O{\empty} m}{
  \ensuremath{\mathcal{A}^{\ifthenelse{\equal{#1}{}}{}{(#1)}}_{#2}(#3)}
}
\DeclareDocumentCommand{\condWtyp}{ O{\empty} O{\empty} m m m}{
  \ensuremath{\mathcal{A}^{\ifthenelse{\equal{#1}{}}{}{(#1)}}_{#2}{(#3 | #4 = #5)}}
}

\DeclareDocumentCommand{\typ}{ O{\empty} O{\empty} m}{
  \ensuremath{\mathcal{T}^{#1}_{[#3]#2}}
}
\DeclareDocumentCommand{\type}{ O{\empty} m}{
  \ensuremath{\mathcal{T}^{#1}_{#2}}
}
\DeclareDocumentCommand{\condTyp}{ O{\empty} O{\empty} m m m}{
  \ensuremath{\mathcal{T}^{#1}_{[#3|#4]#2}(#5)}
}
\DeclareDocumentCommand{\condType}{ O{\empty} m m m}{
  \ensuremath{\mathcal{T}^{#1}_{#2|#3}(#4)}
}

\DeclareDocumentCommand{\ent}{ O{\empty} m}{
  \ensuremath{\mathrm{H}
    \ifthenelse{\equal{#1}{big}}{\big(#2\big)}{
      \ifthenelse{\equal{#1}{Big}}{\Big(#2\Big)}{
        \ifthenelse{\equal{#1}{bigg}}{\bigg(#2\bigg)}{
          \ifthenelse{\equal{#1}{Bigg}}{\Bigg(#2\Bigg)}{
            \ifthenelse{\equal{#1}{normal}}{(#2)}{\left(#2\right)}
          }
        }
      }
    }
  }
}

\DeclareDocumentCommand{\entRate}{ O{\empty} m}{
  \ensuremath{\ol{\mathrm{H}}
    \ifthenelse{\equal{#1}{big}}{\big(#2\big)}{
      \ifthenelse{\equal{#1}{Big}}{\Big(#2\Big)}{
        \ifthenelse{\equal{#1}{bigg}}{\bigg(#2\bigg)}{
          \ifthenelse{\equal{#1}{Bigg}}{\Bigg(#2\Bigg)}{
            \ifthenelse{\equal{#1}{normal}}{(#2)}{\left(#2\right)}
          }
        }
      }
    }
  }
}

\DeclareDocumentCommand{\entPhi}{ O{\empty} m}{
  \ensuremath{\mathrm{H}_{\phi}{}%
    \ifthenelse{\equal{#1}{big}}{\big(#2\big)}{
      \ifthenelse{\equal{#1}{Big}}{\Big(#2\Big)}{
        \ifthenelse{\equal{#1}{bigg}}{\bigg(#2\bigg)}{
          \ifthenelse{\equal{#1}{Bigg}}{\Bigg(#2\Bigg)}{
            \ifthenelse{\equal{#1}{normal}}{(#2)}{\left(#2\right)}
          }
        }
      }
    }
  }
}

\DeclareDocumentCommand{\condEnt}{ O{\empty} m m}{
  \ensuremath{\mathrm{H}
    \ifthenelse{\equal{#1}{big}}{\big({#2\big| #3}\big)}{
      \ifthenelse{\equal{#1}{Big}}{\Big({#2\Big| #3}\Big)}{
        \ifthenelse{\equal{#1}{bigg}}{\bigg({#2\bigg| #3}\bigg)}{
          \ifthenelse{\equal{#1}{Bigg}}{\Bigg({#2\Bigg| #3}\Bigg)}{
            \ifthenelse{\equal{#1}{normal}}{({#2| #3})}{\left(#2 \middle| #3\right)}
          }
        }
      }
    }
  }
}

\DeclareDocumentCommand{\dent}{ O{\empty} m}{
  \ensuremath{\mathrm{h}
    \ifthenelse{\equal{#1}{big}}{\big(#2\big)}{
      \ifthenelse{\equal{#1}{Big}}{\Big(#2\Big)}{
        \ifthenelse{\equal{#1}{bigg}}{\bigg(#2\bigg)}{
          \ifthenelse{\equal{#1}{Bigg}}{\Bigg(#2\Bigg)}{
            \ifthenelse{\equal{#1}{normal}}{(#2)}{\left(#2\right)}
          }
        }
      }
    }
  }
}

\DeclareDocumentCommand{\dentRate}{ O{\empty} m}{
  \ensuremath{\ol{\mathrm{h}}
    \ifthenelse{\equal{#1}{big}}{\big(#2\big)}{
      \ifthenelse{\equal{#1}{Big}}{\Big(#2\Big)}{
        \ifthenelse{\equal{#1}{bigg}}{\bigg(#2\bigg)}{
          \ifthenelse{\equal{#1}{Bigg}}{\Bigg(#2\Bigg)}{
            \ifthenelse{\equal{#1}{normal}}{(#2)}{\left(#2\right)}
          }
        }
      }
    }
  }
}

\DeclareDocumentCommand{\condDent}{ O{\empty} m m}{
  \ensuremath{\mathrm{h}
    \ifthenelse{\equal{#1}{big}}{\big({#2\big| #3}\big)}{
      \ifthenelse{\equal{#1}{Big}}{\Big({#2\Big| #3}\Big)}{
        \ifthenelse{\equal{#1}{bigg}}{\bigg({#2\bigg| #3}\bigg)}{
          \ifthenelse{\equal{#1}{Bigg}}{\Bigg({#2\Bigg| #3}\Bigg)}{
            \ifthenelse{\equal{#1}{normal}}{({#2| #3})}{\left(#2 \middle| #3\right)}
          }
        }
      }
    }
  }
}

\newcommand{\binEntOp}{\mathrm{H}_2}

\DeclareDocumentCommand{\binEnt}{ O{\empty} m}{
  \ensuremath{\binEntOp{
      \ifthenelse{\equal{#2}{\empty}}{}{
        \ifthenelse{\equal{#1}{big}}{\big({#2}\big)}{
          \ifthenelse{\equal{#1}{Big}}{\Big({#2}\Big)}{
            \ifthenelse{\equal{#1}{bigg}}{\bigg({#2}\bigg)}{
              \ifthenelse{\equal{#1}{Bigg}}{\Bigg({#2}\Bigg)}{
                \ifthenelse{\equal{#1}{normal}}{({#2})}{\left(#2\right)}
              }
            }
          }
        }
      }
    }
  }
}

\DeclareDocumentCommand{\binEntInv}{ O{\empty} m}{
  \ensuremath{\binEntOp^{-1}{
      \ifthenelse{\equal{#2}{\empty}}{}{
        \ifthenelse{\equal{#1}{big}}{\big({#2}\big)}{
          \ifthenelse{\equal{#1}{Big}}{\Big({#2}\Big)}{
            \ifthenelse{\equal{#1}{bigg}}{\bigg({#2}\bigg)}{
              \ifthenelse{\equal{#1}{Bigg}}{\Bigg({#2}\Bigg)}{
                \ifthenelse{\equal{#1}{normal}}{({#2})}{\left(#2\right)}
              }
            }
          }
        }
      }
    }
  }
}

\def\barcirc{\mathrel{\barcirci}}
\def\barcirci{{%
    \setbox0\hbox{\ensuremath{\relbar\!\!\relbar}}%
    \rlap{\hbox to \wd0{\hss\ensuremath{\circ}\hss}}\box0
}}

\newcommand{\mkv}{\ensuremath{\barcirc}}

\newcommand{\bernoulli}[1]{\ensuremath{{\BBB(#1)}}}
\DeclareDocumentCommand{\uniform}{ O{\empty} m}{
  \ensuremath{\mathcal{U}
    \ifthenelse{\equal{#1}{big}}{\big({#2}\big)}{
      \ifthenelse{\equal{#1}{Big}}{\Big({#2}\Big)}{
        \ifthenelse{\equal{#1}{bigg}}{\bigg({#2}\bigg)}{
          \ifthenelse{\equal{#1}{Bigg}}{\Bigg({#2}\Bigg)}{
            \ifthenelse{\equal{#1}{normal}}{({#2})}{\left(#2\right)}
          }
        }
      }
    }
  }
}

\newcommand{\DSBS}[1]{\ensuremath{\mathrm{DSBS}(#1)}}

\newcommand{\orv}[1]{\ensuremath{\ol{\rv{#1}}}}
\newcommand{\orvt}[1]{\ensuremath{\ol{\rvt{#1}}}}

\newcommand{\set}[1]{\ensuremath{\mathcal{#1}}}

\newcommand{\AAA}{\ensuremath{\set{A}}}
\newcommand{\BBB}{\ensuremath{\set{B}}}
\newcommand{\CCC}{\ensuremath{\set{C}}}

\newcommand{\EEE}{\ensuremath{\set{E}}}

\newcommand{\HHH}{\ensuremath{\set{H}}}
\newcommand{\III}{\ensuremath{\set{I}}}
\newcommand{\JJJ}{\ensuremath{\set{J}}}
\newcommand{\KKK}{\ensuremath{\set{K}}}

\newcommand{\MMM}{\ensuremath{\set{M}}}

\newcommand{\OOO}{\ensuremath{\set{O}}}
\newcommand{\PPP}{\ensuremath{\set{P}}}
\newcommand{\QQQ}{\ensuremath{\set{Q}}}
\newcommand{\RRR}{\ensuremath{\set{R}}}
\newcommand{\SSS}{\ensuremath{\set{S}}}

\newcommand{\UUU}{\ensuremath{\set{U}}}
\newcommand{\VVV}{\ensuremath{\set{V}}}

\newcommand{\XXX}{\ensuremath{\set{X}}}
\newcommand{\YYY}{\ensuremath{\set{Y}}}

\newcommand{\eps}{\ensuremath{\varepsilon}}

\newcommand{\osubeq}{\ensuremath{\mathord{\sqsubseteq}}}
\newcommand{\osub}{\ensuremath{\mathord{\sqsubset}}}
\newcommand{\osup}{\ensuremath{\mathord{\sqsupset}}}
\newcommand{\osupeq}{\ensuremath{\mathord{\sqsupseteq}}}

\newcommand{\PrIn}{\ensuremath{\PPP_*}}

\newcommand{\CB}{\ensuremath{\rv c}}

\newcommand{\RRRin}{\ensuremath{\RRR_{\mathrm{in}}}}
\newcommand{\RRRht}{\ensuremath{\RRR_{\mathrm{HT}}}}

\newcommand{\RRRpr}{\ensuremath{\RRR_{\mathrm{PR}}}}
\newcommand{\RRRib}{\ensuremath{\RRR_{\mathrm{IB}}}}
\newcommand{\RRRmi}{\ensuremath{\RRR_{\mathrm{MI}}}}

\newcommand{\RRRo}{\ensuremath{\RRR_{\mathrm{o}}}}
\newcommand{\RRRi}{\ensuremath{\RRR_{\mathrm{i}}}}

\newcommand{\SSSo}{\ensuremath{\SSS_{\mathrm{o}}}}
\newcommand{\SSSi}{\ensuremath{\SSS_{\mathrm{i}}}}
\newcommand{\SSSb}{\ensuremath{\SSS_{\mathrm{b}}}}

\newcommand{\jzidx}{\ensuremath{{k}}}

\newcommand{\thetafkt}[1][\empty]{\ensuremath{\hat{\theta}(\ifthenelse{\equal{#1}{\empty}}{\rho,\aii,\bii}{#1})}}

\newcommand{\aii}{\ensuremath{\alpha}}

\newcommand{\bii}{\ensuremath{\beta}}

\newcommand{\CST}[1][\empty]{\ensuremath{C_{\ifthenelse{\equal{#1}{\empty}}{\aii,\bii}{#1}}}}

\ifdefined\tikzextpath
\usetikzlibrary{external}
\fi

\pgfplotsset{compat=1.14}

\usetikzlibrary{arrows}
\usetikzlibrary{arrows.meta}
\usetikzlibrary{decorations}
\usetikzlibrary{backgrounds}
\usetikzlibrary{fit}
\usetikzlibrary{calc}
\usetikzlibrary{intersections}
\usetikzlibrary{decorations}
\usetikzlibrary{spy}
\usetikzlibrary{positioning}
\usetikzlibrary{shapes}
\usetikzlibrary{graphs}
\usetikzlibrary{shapes.multipart}
\usetikzlibrary{automata}
\usetikzlibrary{fixedpointarithmetic}
\usetikzlibrary{decorations.markings}

\definecolor{myred}{RGB}{128, 0, 0}
\definecolor{myblue}{RGB}{0, 0, 128}

\tikzstyle{block} = [draw,fill=RoyalBlue!30,minimum size=2em,rounded corners=2mm]
\tikzstyle{symb}=[]
\def\radius{.8mm} 
\tikzstyle{branch}=[fill,shape=circle,minimum size=3pt,inner sep=0pt]
\tikzstyle{s}=[shift={(0mm,\radius)}]

\makeatletter
\newcommand{\includetikz}[1]{%
  \ifdefined\tikzexternalize%
  \filename@parse{#1}%
  \tikzsetnextfilename{\filename@base}%
  \fi%
  \input{#1.tikz}%
}
\makeatother

\AtBeginDocument{
  \label{CorrectFirstPageLabel}
  \def\fpage{\pageref{CorrectFirstPageLabel}}
}

\allowdisplaybreaks
\usepackage[cal=cm,
calscaled=.94,
bb=ams,
frak=euler,
frakscaled=.97,
scr=rsfs]{mathalpha}

\pgfplotsset{
        table/search path={data/bin_region/,data/bounds/data},
    }

\setcitestyle{numbers}
\makeatletter

\let\c@conjecture\relax
\makeatother
\newtheorem{conjecture}[theorem]{\sc Conjecture}

\renewcommand{\figurename}{Figure}
\renewcommand{\tablename}{Table}
\crefname{assumption}{Assumption}{Assumptions}
\Crefname{assumption}{Assumption}{Assumptions}
\crefname{def}{Definition}{Definitions}
\Crefname{def}{Definition}{Definitions}
\crefname{pro}{Proposition}{Propositions}
\Crefname{pro}{Proposition}{Propositions}

\renewcommand{\binEntOp}{\mathrm{h}_2}

\DeclareDocumentCommand{\Ntoo}{ O{\empty} m}{
  \ensuremath{
    \ifthenelse{\equal{#1}{big}}{\big[1\,{:}\,#2\big]}{
      \ifthenelse{\equal{#1}{Big}}{\Big[1\,{:}\,#2\Big]}{
        \ifthenelse{\equal{#1}{bigg}}{\bigg[1\,{:}\,#2\bigg]}{
          \ifthenelse{\equal{#1}{Bigg}}{\Bigg[1\,{:}\,#2\Bigg]}{
            \ifthenelse{\equal{#1}{normal}}{[1\,{:}\,#2]}{\left[1\,{:}\,#2\right]}
          }
        }
      }
    }
  }
}

\RenewDocumentCommand{\uniform}{ O{\empty} m}{
  \ensuremath{\mathfrak{U}
    \ifthenelse{\equal{#1}{big}}{\big({#2}\big)}{
      \ifthenelse{\equal{#1}{Big}}{\Big({#2}\Big)}{
        \ifthenelse{\equal{#1}{bigg}}{\bigg({#2}\bigg)}{
          \ifthenelse{\equal{#1}{Bigg}}{\Bigg({#2}\Bigg)}{
            \ifthenelse{\equal{#1}{normal}}{({#2})}{\left(#2\right)}
          }
        }
      }
    }
  }
}
\RenewDocumentCommand{\bernoulli}{ O{\empty} m}{
  \ensuremath{\mathfrak{B}
    \ifthenelse{\equal{#1}{big}}{\big({#2}\big)}{
      \ifthenelse{\equal{#1}{Big}}{\Big({#2}\Big)}{
        \ifthenelse{\equal{#1}{bigg}}{\bigg({#2}\bigg)}{
          \ifthenelse{\equal{#1}{Bigg}}{\Bigg({#2}\Bigg)}{
            \ifthenelse{\equal{#1}{normal}}{({#2})}{\left(#2\right)}
          }
        }
      }
    }
  }
}

\newcommand{\shiftleft}{\hspace*{-.2\textwidth}}
\newcommand{\shiftleftB}{\hspace*{-.13\textwidth}}
\newcommand{\shiftleftC}{\shiftleft}

\makeatletter
\newtoks\@tabtoks
\newcommand\addtabtoks[1]{\global\@tabtoks\expandafter{\the\@tabtoks#1}}
\newcommand\eaddtabtoks[1]{\edef\mytmp{#1}\expandafter\addtabtoks\expandafter{\mytmp}}
\newcommand*\resettabtoks{\global\@tabtoks{}}
\newcommand*\printtabtoks{\the\@tabtoks}
\makeatother

\pgfmathsetmacro{\largestGap}{0.00019856728373242833805660534363}
\pgfmathsetmacro{\rateLGap}{0.67567613908269752442947719828226}
\pgfmathsetmacro{\muLGap}{0.35972009086758860441079832526157}
\pgfmathsetmacro{\myp}{0.10000000000000000555111512312578}
\def\Plgap{{
{
  {  {0.99535814621735340652500000000000, 0.0024976755984442331907500000000000,  }, {0.0024983440039573106433250000000000, 0.00034313919194834475000000000000000,  }, },
  {  {0.0021426038576540942750000000000000, 0.99500307447656326760925000000000,  }, {0.00000090592103518855667500000000000000, 0.0021561107330441544500000000000000,  }, },
},
{
  {  {0.0021426038576540942750000000000000, 0.0000015743265482660092500000000000000,  }, {0.99500240607105019015667500000000, 0.0021561107330441544500000000000000,  }, },
  {  {0.00035664606733840492500000000000000, 0.0024976755984442331907500000000000,  }, {0.0024983440039573106433250000000000, 0.99534463934196334635000000000000,  }, },
},
}}
\pgfmathsetmacro{\mya}{0.80000000000000004440892098500626}
\pgfmathsetmacro{\myabsc}{0.07658027968814740127356799348490}
\pgfmathsetmacro{\myIUV}{0.29110316603236779942065481918689}
\pgfmathsetmacro{\myR}{0.42281045524016253045473945348931}
\pgfmathsetmacro{\myIUVbsc}{0.28559362818656935933958607165550}
\pgfmathsetmacro{\myGap}{0.00550953784579845309149481735744}

\RenewDocumentCommand{\typ}{ O{\empty} O{\empty} m}{
  \ensuremath{\mathcal{T}_{#2}(#3)}
}
\DeclareDocumentCommand{\condTyp}{ O{\empty} O{\empty} m m m}{
  \ensuremath{\mathcal{T}_{#2}(#3|#4=#5)}
}
\def\jzidx{j}

\makeatletter
\newenvironment{proof*}[1][\proofname]{\par
  \vspace{\topsep}%
  \pushQED{\qed}%
  \normalfont \partopsep=\z@skip \topsep=\z@skip
  \trivlist
  \item[\hskip\labelsep
        \itshape
    #1\@addpunct{.}]\ignorespaces
}{%
  \popQED\endtrivlist\@endpefalse
}
\makeatother

\begin{document}
\def\mytitle{Distributed information-theoretic clustering}
\title{\mytitle}
\shorttitle{\mytitle}
\shortauthorlist{G.~Pichler, P.~Piantanida, and~G.~Matz}

\author{{    \sc Georg~Pichler}$^*$,\\[2pt]
  Institute of Telecommunications, TU Wien, Vienna, Austria.\\
  $^*${\email{Corresponding author: georg.pichler@ieee.org}}\\[2pt]
    {\sc Pablo~Piantanida}\\[2pt]
  Universit\'e Paris-Saclay, CNRS, CentraleSup\'elec,  Laboratoire des Signaux et Syst\`emes,  Gif-sur-Yvette, France and Montreal Institute for Learning Algorithms (Mila), Universit\'e de Montr\'eal, QC, Canada.\\
  {pablo.piantanida@centralesupelec.fr}\\[6pt]
    {\sc and}\\[6pt]
    {\sc Gerald~Matz} \\[2pt]
  Institute of Telecommunications, TU Wien, Vienna, Austria.\\
  {gerald.matz@tuwien.ac.at}
  }

\maketitle

\begin{abstract}
  {We study a novel multi-terminal source coding setup motivated by the biclustering problem. Two separate encoders observe two \iid sequences $\rv x^n$ and $\rv y^n$, respectively.
The goal is to find rate-limited encodings $f(x^n)$ and $g(z^n)$ that maximize the mutual information $\mutInf[big]{f(\rv x^n)}{g(\rv y^n)}/n$.
We discuss connections of this problem with hypothesis testing against independence, pattern recognition, and the information bottleneck method.
Improving previous cardinality bounds for the inner and outer bounds allows us to thoroughly study the special case of a binary symmetric source and to quantify the gap between the inner and the outer bound in this special case.
Furthermore, we investigate a multiple description (MD) extension of the Chief Operating Officer (CEO) problem with mutual information constraint. Surprisingly, this MD-CEO problem permits a tight single-letter characterization of the achievable region.
}
  {source coding; mutual information; information bottleneck; CEO problem}
\end{abstract}

\section{Introduction}
\label{sec:introduction}

The recent decades witnessed a rapid proliferation of digital data in a myriad of repositories such as internet fora, blogs, web applications, news, emails and the social media bandwagon. A significant part of this data is unstructured and it is thus hard to extract relevant information. This results in a growing need for a fundamental understanding and efficient methods for analyzing data and discovering valuable and relevant knowledge from it in the form of structured information. 

When specifying certain hidden (unobserved) features of interest, the problem then consists of extracting those relevant features from a measurement, while neglecting other, irrelevant features.
Formulating these idea in terms of lossy source compression~\cite{Shannon1993CodingTheoremsForADiscreteSourceWithAFidelityCriterion}, we can quantify the complexity of the representations via its compression rate and the fidelity via the information provided about specific (unobserved) features.

In this paper, we introduce and study the distributed clustering problem from a formal information-theoretic perspective. Given correlated samples $\rv x_1, \rv x_2$ observed at two different encoders, the aim is to extract a description from each sample such that the descriptions are maximally informative about each other. In other words, each encoder tries to find a (lossy) description $\rv w_j = f_j(\rv x_j^n)$ of its observation $\rv x_j^n$  subject to a complexity requirement (coding rate), maximizing the mutual information $\mutInf{\rv w_1}{\rv w_2}$. Our goal is to characterize the optimal tradeoff between the \emph{relevance} (mutual information between the descriptions) and the \emph{complexity} of those descriptions (encoding rate).

\subsection{Related work}
\label{sec:related}

\emph{Biclustering} (or \emph{co-clustering}) was first explicitly considered by Hartigan~\cite{Hartigan1972Direct} in 1972.
A historical overview of biclustering including additional background can be found in \cite[Section~3.2.4]{Mirkin1996Mathematical}.
In general, given an $S \times T$ data matrix $(a_{st})$, the goal of a biclustering algorithm~\cite{Madeira2004Biclustering} is to find partitions $\BBB_k \subseteq \{1, \dots, S\}$ and $\CCC_l \subseteq \{1, \dots, T\}$, $k=1 \dots K$, $l=1 \dots L$ such that all the elements of the `biclusters' $(a_{st})_{s \in \BBB_k, t \in \CCC_l}$ are homogeneous in a certain sense. 
The measure of homogeneity of the biclusters depends on the specific application.
The method received renewed attention when Cheng and Church~\cite{Cheng2000Biclustering} applied it to gene expression data. Many biclustering algorithms have been developed since (\eg, see~\cite{Tanay2005Biclustering} and the references therein). An introductory overview of clustering algorithms for gene expression data can be found in the lecture notes~\cite{Shar2006Analysis}. The \emph{information bottleneck} (IB) method, which can be viewed as a 
uni-directional information-theoretic variant of biclustering, was successfully applied to gene expression data as well~\cite{Slonim2005Information}.

In 2003, Dhillon \etal~\cite{Dhillon2003Information} adopted an information-theoretic approach to biclustering. They used mutual information to characterize the quality of a biclustering. Specifically, for the special case when the underlying matrix represents the joint probability distribution of two discrete random variables $\rv x$ and $\rv y$, \ie, $a_{st} = \Prob{\rv x=s, \rv y=t}$, their goal was to find clustering functions $f \colon \{1,\dots,S\} \to \{1,\dots,K\}$ and $g \colon \{1,\dots,T\} \to \{1,\dots,L\}$ that maximize $\mutInf[big]{f(\rv x)}{g(\rv y)}$ for specific $K$ and $L$.
This idea was successfully employed in numerous research papers since, \eg, \cite{Gokcay2002Information,Mueller2012Information,Steeg2014Demystifying,Kraskov2009MIC}, where mutual information is typically estimated from samples.

In the present work, we investigate a theoretical extension of the approach in~\cite{Dhillon2003Information}, where we consider blocks of $n$ \iid sources and $S_n$, $K_n$, $T_n$, and $L_n$ scale exponentially in the blocklength $n$.
The resulting \emph{information-theoretic biclustering problem} turns out to be equivalent to hypothesis testing against independence with multi-terminal data compression~\cite{Han1987Hypothesis} and to a pattern recognition problem~\cite{Westover2008Achievable}. Both these problems are not yet solved in general (for a survey on the hypothesis testing problem, see~\cite{Han1998Statistical}). The pattern recognition problem has been extensively studied on doubly symmetric binary and jointly Gaussian sources.

A special case of the information-theoretic biclustering problem is given by the \emph{IB problem}, studied in \cite{Gilad2003Information}, based on the IB method~\cite{Tishby2000Information}. This problem is solved in terms of a single-letter characterization and is known to be equivalent to source coding under logarithmic loss. A generalization to multiple terminals, the CEO problem under logarithmic loss~\cite{Courtade2014Multiterminal}, is currently only solved under specific Markov constraints. 

\subsection{Contributions}
\label{sec:contribution}

The aim of the this paper is to characterize the achievable region of the information-theoretic biclustering problem, its extensions and special cases, and connect them to known problems in network information theory.
This problem is fundamentally different from `classical' distributed source coding problems like distributed lossy compression~\cite[Chapter~12]{ElGamal2011Network}. Usually, one aims at reducing redundant information, \ie, information that is transmitted by multiple encoders, as much as possible, while still guaranteeing correct decoding. By contrast, in the biclustering problem we are interested in maximizing this very redundancy. In this sense, it is complementary to conventional distributed source coding and requires adapted proof techniques.

More specifically, the main contributions are as follows.
\begin{itemize}
\item 
We formally prove the equivalence of the information-theoretic biclustering, the hypothesis testing~\cite{Han1987Hypothesis}, and the pattern recognition problem~\cite{Westover2008Achievable} (\cref{thm:equiv}) and connect it to the IB problem~\cite{Tishby2000Information,Gilad2003Information} (\cref{pro:equiv}).
\item 
We extensively study the doubly symmetric binary source (DSBS) as a special case (\cref{sec:binary}).
In order to perform this analysis, we require stronger cardinality bounds
than the ones usually obtained using the convex cover method~\cite[Appendix~C]{ElGamal2011Network}.
\item 
We are able to improve upon the state-of-the-art cardinality bounding techniques by combining the convex cover method with the perturbation method~\cite[Appendix~C]{ElGamal2011Network} and leveraging ideas similar to~\cite{Nair2013Upper}, which allow us to restrict our attention to the extreme points of the achievable region.
The resulting bounds (\cref{pro:cardinality_bound})
allow for the use of binary auxiliary random variables in the case of binary sources.
\item 
We show that \cite[Conjecture~1]{Westover2008Achievable} does not hold
(\cref{pro:counterexample-conj1,pro:binary-counterexample}).
\item 
Based on a weaker conjecture
(\cref{conj:binary}),
we argue that there is indeed a gap between the outer and the inner bound for a DSBS.
(\cref{conj:loose_bound2}).
\item 
We propose an extension of the CEO problem under an information constraint, studied in \cite{Courtade2014Multiterminal}, which 
requires \emph{multiple description} (MD) coding~\cite{ElGamal1982Achievable} (see~\cite{Goyal2001Multiple} for applications) to account for the possibility that descriptions are not delivered.
Using tools from submodularity theory and convex analysis, we are able to provide a complete single-letter characterization of the resulting achievable region
(\cref{thm:mdceo:main}),
which has the remarkable feature that it allows to exploit rate that is in general insufficient for successful typicality decoding.
\end{itemize}

\subsection{Notation and conventions}
\label{sec:notation}

For a total order $\osub$ on a set $\EEE$ (\cf \cite[Definition~1.5]{Rudin1976Principles}) and $e \in \EEE$ we will use the notation $\osup e \defas \{e\prm \in \EEE : e\prm \sqsupset e\}$ and accordingly for $\osupeq$, $\osub$ and $\osubeq$.
\Eg, given the total order $\osub$ on $\{1,2,3\}$ with $3 \sqsubset 1 \sqsubset 2$, we have $\osup 3 = \{1,2\}$, $\osup 1 = \{2\}$ and $\osup 2 = \varnothing$.

We will use the shorthand $[l\,{:}\,k] \defas \{l,l+1,\dots,k-1,k\}$.
The notation $\ind{\AAA}{}$, $\ol{\AAA}$, $\cvx{\AAA}$, and $\card{\AAA}$ is used for the indicator, topological closure, convex hull, and cardinality of a set $\AAA$, respectively.
When there is no possibility of confusion we identify singleton set with its element, \eg, we write $\{1,2,3\} \setminus 1 = \{2,3\}$.
Let $\RR_+$and $\RR_-$ be the set of non-negative and non-positive reals, respectively.

We denote random quantities and their realizations by capital and lowercase letters, respectively. Furthermore, vectors are indicated by bold-face type and have length $n$, if not otherwise specified.
Random variables are assumed to be supported on finite sets and unless otherwise specified, the same letter is used for the random variable and for its support set, \eg, $\rv y$ takes values in $\YYY$ and $\rv x_3$ takes values in $\XXX_3$.
Given a random variable $\rv x$, we write $\p[\rv x]{}$ for its probability mass function (pmf), where the subscript might be omitted if there is no ambiguity.
The notation $\rv x \sim \p{}$ indicates that $\rv x$ is distributed according to $\p{}$ and $\rv x \sim \bernoulli{p}$ and $\rv y \sim \uniform{\YYY}$ denote a Bernoulli distributed random variable $\rv x$ with parameter $p \in [0, 1]$ and a uniformly distributed random variable $\rv y$ on its support set $\YYY$.
We use $\Exp{\rv x}$ and $\Prob{\AAA}$ for the expectation of the random variable $\rv x$ and the probability of an event $\AAA$, respectively.
Subscripts indicate parts of vectors, \eg, $\vt x_\AAA \defas (x_i)_{i \in \AAA}$ for a vector $\vt x = (x_1, x_2, \dots, x_n)$ and $\AAA \subseteq \Ntoo{n}$.
We further use the common notation $\vt x_i^j \defas \vt x_{\{i,\dots,j\}}$, $\vt x^{j} \defas \vt x_1^j$.
If a vector is already carrying a subscript, it will be separated by a comma, \eg, $\vt x_{3,1}^5 = (\vt x_3)_1^5 = (\vt x_3)^5$.
Let $\vt 0$ denote the all-zeros vector and $\vt e_i = (e_{i,1}, e_{i,2}, \dots, e_{i,n}) \in \RR^n$ the $i$th canonical base vector, \ie, $e_{i,j} = \ind{i}{j}$.
We use the notation of~\cite[Chapter~2]{Cover2006Elements} for information-theoretic quantities.
In particular, given random variables $(\rv x, \rv y, \rv z)$ and pmfs $\p{}$ and $\q{}$, $\ent{\rv x}$, $\condEnt{\rv x}{\rv y}$, $\mutInf{\rv x}{\rv y}$, $\condMutInf{\rv x}{\rv y}{\rv z}$, and $\DKL{\p{}}{\q{}}$ denote entropy, conditional entropy, mutual information, conditional mutual information, and Kullback-Leibler divergence, respectively.
All logarithms in this paper are to base $\ee{}$ and therefore all information theoretic quantities are measured in nats.
The notation $\binEnt{p} \defas -p \log p - (1-p) \log(1-p)$ is used for the binary entropy function, $a \ast b \defas a(1-b)+(1-a)b$ is the binary convolution operation and the symbol $\oplus$ denotes binary addition.
The notation $\rv x \mkv \rv y \mkv \rv z$ indicates that
$\rv x$, $\rv y$, and $\rv z$ form a Markov chain in this order and $\rv x \perp \rv y$ denotes that $\rv x$ and $\rv y$ are independent random variables.
Slightly abusing notation we consider $\varnothing$ to be a degenerate random variable that is almost surely a constant, such that, \eg, $\rv x \mkv \varnothing \mkv \rv y \Leftrightarrow \rv x \perp \rv y$.
To simplify the presentation (\cf~\cite{ElGamal2011Network}) when generating codebooks, we will assume that the codebook size is an integer.
We will use superscript to indicate that a relation follows from a specific equation. \Eg, the inequality $a \stackrel{\scriptsize (42)}{\le} b$ follows from equation $(42)$.

\section{Problem statement}
\label{sec:problem}

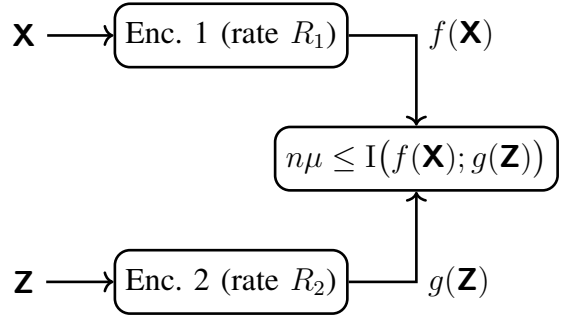
\begin{figure}[tb]
  \centering
  \begin{tikzpicture}[row sep=12mm,
  column sep=9mm,
  ampersand replacement=\&,
  line width=1pt,
  baseline=2mm,
  block/.style={draw, rectangle, minimum height=2em, rounded corners=2mm},
  dec/.style={draw, rectangle, minimum height=2em, rounded corners=2mm},
  symb/.style={},
  ]  \matrix{
        \node [symb] (x) {$\rvt x$}; \& 
    \node [block] (enc1) {Enc.~1 (rate $R_1$)}; \&
    \&
    \\
    \& \&
    \coordinate (dummy1); 
    \\
    \node [symb] (z) {$\rvt y$}; \&
    \node [block] (enc2) {Enc.~2 (rate $R_2$)}; \&
    \&
    \\
  };
  \node [dec,minimum width=3.5cm,below=0pt of dummy1,anchor=center] (dec) {$n\mu \le \mutInf[big]{f(\rvt x)}{g(\rvt y)}$};

  \draw [->] (x) -- (enc1);
  \draw [->] (z) -- (enc2);

    \draw [->] (enc1) -| node[symb,right] (m1) {$f(\rvt x)$} (dec);
  \draw [->] (enc2) -| node[symb,right] (m2) {$g(\rvt y)$} (dec);
\end{tikzpicture}
  \caption{Biclustering of two memoryless sources.}
  \label{fig:biclustering_2}
\end{figure}

In this~\lcnamecref{sec:problem} we will introduce the \emph{information-theoretic biclustering problem} (or \emph{biclustering problem} for short) with two sources and provide bounds for its achievable region.
A schematic overview of the problem is presented in~\cref{fig:biclustering_2}.
Let $(\rv x, \rv y)$ be two random variables. The random vectors $(\rvt x, \rvt y)$ consist of $n$ \iid copies of $(\rv x, \rv y)$. 
Given a block length $n \in \NN$ and coding rates $R_1, R_2 \in \RR_+$, an $(n,R_1,R_2)$-code consists of two functions $f\colon \XXX^n \rightarrow \MMM_1$ and $g\colon \YYY^n \rightarrow \MMM_2$ 
such that the finite sets $\MMM_k$ satisfy $\log |\MMM_k| \le nR_k$, $k \in \{1,2\}$.
Thus, the coding rates $R_1$ and $R_2$ limit the complexity of the encoders.
In contrast to rate-distortion theory, we do not require a specific distortion measure; 
rather, we quantify the quality of a code in pure information-theoretic terms, namely via mutual information. The idea is to find functions $f$ and $g$ that extract 
a compressed version of the common randomness in the observed data $\rvt x$ and $\rvt y$. To this end, we use the normalized mutual information $\mutInf[big]{f(\rvt x)}{g(\rvt y)}/n$ to quantify the relevance of the two encodings.

\begin{definition}
  \label[def]{def:biclustering:achievability}
  A triple $(\mu, R_1, R_2)$ is \emph{achievable} if, for some $n \in \NN$, there exists an $(n,R_1,R_2)$-code $(f,g)$ such that 
  \begin{align}
    \label{eq:biclustering:achievability}
    \frac1n \mutInf[big]{f(\rvt x)}{g(\rvt y)} &\ge \mu .
  \end{align}
  The achievable region $\ol{\RRR}$ is defined as the
  closure of the set $\RRR$ of achievable triples.
\end{definition}

\begin{remark}
  Note that a standard time-sharing argument shows that $\ol{\RRR}$ is a convex set (\cf \cite[Section~4.4]{ElGamal2011Network}).
\end{remark}

\noindent
We also point out that stochastic encodings cannot enlarge the achievable region as any stochastic encoding can be represented as the convex combination of deterministic encodings and $\ol{\RRR}$ is convex.

\section{Equivalent problems}
\label{sec:equivalent-problems}

The biclustering problem turns out to be equivalent to a hypothesis testing and a pattern recognition problem. In this \lcnamecref{sec:equivalent-problems} we will clarify this equivalence by showing that the multi-letter regions agree. These equivalences will provide us with the achievability of $\RRR_*$, the `multi-letter' region $\RRR_*$ of the biclustering problem.

\begin{definition}
  \label[def]{def:multi-letter}
  Let $\RRR_*$ be the set of triples $(\mu, R_1, R_2)$ such that there exist $n \in \NN$ and random variables $\rv u$, $\rv v$ satisfying $\rv u \mkv \rvt x \mkv \rvt y \mkv \rv v$ and
  \begin{align}
    nR_1 &\ge \mutInf{\rv u}{\rvt x} , \label{eq:multi-letter:1}\\*
    nR_2 &\ge \mutInf{\rv v}{\rvt y} , \label{eq:multi-letter:2}\\*
    n\mu &\le \mutInf{\rv u}{\rv v} . \label{eq:multi-letter:mu}
  \end{align}
\end{definition}

Next, we consider the hypothesis testing problem with data compression when testing against independence \cite[Section~6]{Han1987Hypothesis} and the pattern recognition problem~\cite{Westover2008Achievable}.
For completeness sake we briefly describe the problem setups.

\begin{definition}[Hypothesis testing against independence]
  Given the potentially dependent sources $(\rv x, \rv y)$, define the independent random variables $(\orv x, \orv y) \sim \p[\rv x]{} \times \p[\rv y]{}$.
  An $(n,R_1,R_2)$ hypothesis test consists of an $(n,R_1,R_2)$-code $(f_n,g_n)$ and a set $\AAA_n \subseteq \MMM_1 \times \MMM_2$, where $\MMM_1$ and $\MMM_2$ are the ranges of $f_n$ and $g_n$, respectively. The type I and type II error probabilities of $(f_n,g_n,\AAA_n)$ are defined as $\alpha_n \defas \Prob{\big(f_n(\rvt x), g_n(\rvt y)\big) \in \AAA_n}$ and $\beta_n \defas \Prob{\big(f_n(\orvt x), g_n(\orvt y)\big) \notin \AAA_n}$, respectively. A triple $(\mu,R_1,R_2)$ is \emph{HT-achievable} if, for every $\eps > 0$, there is a sequence of $(n,R_1,R_2)$ hypothesis tests $(f_n,g_n,\AAA_n)$, $n \in \NN$ such that 
  \begin{align}
    \lim_{n\to\infty} \alpha_n &\le \eps , \label{eq:hypo:type1} \\*
    \lim_{n\to\infty} - \frac 1n \log\beta_n &\ge \mu . \label{eq:hypo:type2}
  \end{align}
  Let $\RRRht$ denote the set of all HT-achievable triples.
\end{definition}

\begin{definition}[Pattern recognition]
  \label[def]{def:pr}
  Let $\big(\rvt x(i), \rvt y(i)\big)$ be $n$ \iid copies of $(\rv x, \rv y)$, independently generated for each $i \in \NN$.
  A triple $(\mu, R_1, R_2)$ is said to be \emph{PR-achievable} if, for any $\eps > 0$, there is some $n \in \NN$, such that there exists an $(n, R_1, R_2)$-code $(f,g)$ and a function $\phi \colon (\MMM_1)^{\ee{n\mu}} \times \MMM_2 \to \Ntoo{\ee{n\mu}}$ with
  \begin{align}
    \Prob{\rv w = \phi\big(\CB, g(\rvt y(\rv w))\big)} \ge 1-\eps , \label{eq:pattern_condition}
  \end{align}
  where $\CB \defas f(\rvt x(i))_{i \in \Ntoo{\ee{n\mu}}}$ is the compressed codebook and $\big(\rvt x(i), \rvt y(i)\big)_{i \in \NN} \perp \rv w \sim \uniform{\Ntoo{\ee{n\mu}}}$.
  Let $\RRRpr$ denote the set of all PR-achievable triples.
\end{definition}

\begin{remark}
  \label{rmk:westover-error}
  The variant of the inner bound for the pattern recognition problem stated in \cite[Theorem~1]{Westover2008Achievable} is flawed. To see this, note that (using the notation of \cite{Westover2008Achievable}) the point $(R_x=0, R_y=b, R_c=b)$ is contained in $\RRRin$ (choose $U = V = \varnothing$) for any $b>0$ even if the random variables $X$ and $Y$ are independent. But this point is clearly not achievable in general.
  However, the region $\RRRin$ defined in the right column of \cite[p.~303]{Westover2008Achievable} coincides with our findings and the proof given in \cite[Appendix~A]{Westover2008Achievable} holds for this region.
\end{remark}

The biclustering, hypothesis testing and pattern recognition problems are equivalent in the sense that their `multi-letter' regions agree. The proof of this result is given in \cref{apx:proof:equiv}.
\begin{theorem}
  \label{thm:equiv}
  $\ol{\RRR} = \ol{\RRR_*} = \ol{\RRRht} = \ol{\RRRpr}$.
\end{theorem}

\section{Bounds on the achievable region}
\label{sec:bounds_twosources}

The following inner and outer bound on the achievable region follow from the corresponding results on the hypothesis testing and pattern recognition problems.

\begin{theorem} 
  \label{thm:inner-bound}
  We have $\RRRi \subseteq \ol{\RRR}$ where 
  \begin{align}
    \label{eq:achiev}
    \RRRi \defas \bigcup_{\rv u, \rv v} \big\{ (\mu,R_1,R_2):\; 
    R_1 \ge \mutInf{\rv u}{\rv x},\;
    R_2 \ge \mutInf{\rv v}{\rv y},\;
    \mu \le \mutInf{\rv u}{\rv v}\big\},
  \end{align}
with auxiliary random variables $\rv u$, $\rv v$ satisfying $\rv u \mkv \rv x \mkv \rv y \mkv \rv v$.
\end{theorem}

\begin{theorem}
  \label{thm:outer-bound}
  We have   $\RRR \subseteq \RRRo$, where
  \begin{align}
\RRRo  \defas \smash{\bigcup_{\rv u, \rv v}} \big\{ (\mu,R_1,R_2) :  
	R_1 &\ge \mutInf{\rv u}{\rv x},\;
	R_2 \ge \mutInf{\rv v}{\rv y},
	\eqnl \mu &\le \mutInf{\rv u}{\rv x} + \mutInf{\rv v}{\rv y} - \mutInf{\rv u \rv v}{\rv x \rv y} 
  \big\} , \label{eq:outer-bound:Ro} \end{align}
with $\rv u$ and $\rv v$ any pair of random variables satisfying $\rv u \mkv \rv x \mkv \rv y$ and $\rv x \mkv \rv y \mkv \rv v$.
\end{theorem}
Using \cref{thm:equiv}, \cref{thm:inner-bound} follows either from~\cite[Corollary~6]{Han1987Hypothesis}, or from~\cite[Appendix~A]{Westover2008Achievable}.
\Cref{thm:outer-bound} follows from~\cite[Appendix~B]{Westover2008Achievable} using~\cref{thm:equiv}.

The main differences between the outer and the inner bound lie in the Markov conditions, a phenomenon that also occurs with Berger--Tung type bounds (\cf \cite[Chapter~7]{Tung1978Multiterminal} or \cite[Section~12.2]{ElGamal2011Network}).
Note that $\RRRo$ and $\RRRi$ would coincide if the Markov condition $\rv u \mkv \rv x \mkv \rv y \mkv \rv v$ were imposed in the definition of $\RRRo$.
The region $\RRRo$ is convex since a time-sharing variable can be 
incorporated into $\rv u$ and $\rv v$.
The inner bound $\RRRi$, however, can be improved by convexification.

Numerical evaluation of $\RRRo$ and $\RRRi$ requires the cardinalities of the auxiliary random variables to be bounded. We therefore complement \cref{thm:inner-bound,thm:outer-bound} with the following result, whose proof is provided in~\cref{apx:proof:cardinality_bound}.

\begin{proposition}
  \label[pro]{pro:cardinality_bound}
  Let $\SSSo$ and $\SSSi$ be defined like $\RRRo$ and $\RRRi$, respectively, but with the additional cardinality bounds $\card{\UUU} \le \card{\XXX}$ and $\card{\VVV} \le \card{\YYY}$.
  We then have $\cvx{\SSSo} = \RRRo$ and $\cvx{\SSSi} = \cvx{\RRRi} \subseteq \ol{\RRR}$.
\end{proposition}

\begin{remark}
  Note that $\cvx{\SSSi}$ can be explicitly expressed as
    \begin{align}\label{eq:cardinality_bound}
    \cvx{\SSSi} = \bigcup_{\rv u, \rv v, \rv q}
    \big\{ (\mu,R_1,R_2):\; 
    R_1 \ge \condMutInf{\rv u}{\rv x}{\rv q},\;
    R_2 \ge \condMutInf{\rv v}{\rv y}{\rv q},\;
    \mu \le \condMutInf{\rv u}{\rv v}{\rv q} 
    \big\} ,
  \end{align}
  where $\rv u$, $\rv v$, and $\rv q$ are random variables such that 
  $\p[\rv x, \rv y, \rv u, \rv v, \rv q]{} = \p[\rv q]{}\, \p[\rv x, \rv y]{}\,\pcond[\rv u][\rv x, \rv q]{}{}\, \pcond[\rv v][\rv y, \rv q]{}{}$,
    $\card{\UUU} \le \card{\XXX}$, $\card{\VVV} \le \card{\YYY}$, and $\card{\QQQ} \le 3$.

  The cardinality bound $\card{\QQQ} \le 3$ follows directly from the strengthened Carathéodory theorem~\cite[Theorem~18(ii)]{Eggleston1958Convexity} because $\cvx{\RRRi}$ is the convex hull of a connected set in $\RR^3$.
\end{remark}

Note that the cardinality bounds in this result are tighter than 
the usual bounds obtained with the convex cover method~\cite[Appendix~C]{ElGamal2011Network}, where the cardinality has to be increased by one.
We will exploit this fact with binary sources in \cref{sec:binary}, to show that binary auxiliary random variables suffice.
The smaller cardinality bounds come at the cost of convexification for the outer bound since in contrast to $\RRRo$, the region $\SSSo$ is not necessarily convex.

A tight bound on the achievable region can be obtained if $\mu$ is not greater than the Gács-Körner common information (\cf \cite{Gacs1973Common,Wagner2011Distributed,Witsenhausen1975Sequences}) of $\rv x$ and $\rv y$, as stated in the following \lcnamecref{cor:comm-inf}.
\begin{corollary}
  \label{cor:comm-inf}
  If $\rv y = \zeta_1(\rv x) = \zeta_2(\rv y)$ is common to $\rv x$ and $\rv y$ in the sense of~\cite{Wagner2011Distributed} and $0 \le \mu \le \ent{\rv y}$ then $(\mu, R_1, R_2) \in \ol{\RRR}$ if and only if $\mu \le \min \{R_1, R_2\}$.
\end{corollary}
\begin{proof}
  \Cref{thm:outer-bound} entails $\mu \le \min\{R_1, R_2\}$ for any $(\mu, R_1, R_2) \in \ol{\RRR}$.
  With $\rv u = \rv v = \rv y$, \cref{thm:inner-bound} implies $(\ent{\rv y}, \ent{\rv y}, \ent{\rv y}) \in \ol{\RRR}$. Using time-sharing with $\vt 0 \in \ol{\RRR}$ we obtain $(\mu, \mu, \mu) \in \ol{\RRR}$ for $0 \le \mu \le \ent{\rv y}$ and hence $(\mu, R_1, R_2) \in \ol{\RRR}$ if $\mu \le \min\{R_1, R_2\}$.
\end{proof}

\section{Doubly symmetric binary source}
\label{sec:binary}

In this \lcnamecref{sec:binary}, we analyze the achievable region for a DSBS. The same region (\cf \cref{thm:equiv}) was previously analyzed in \cite{Westover2008Achievable} in the context of a pattern recognition problem. We obtain additional results, disproving \cite[Conjecture~1]{Westover2008Achievable}.
In particular, we conjecture that there is a gap between the inner bound $\cvx{\SSSi}$ and the outer bound $\RRRo$ for the DSBS. To support this conjecture, we analyze a region $\SSSb$, previously introduced by the authors of \cite{Westover2008Achievable}, with the property that $\SSSb \subseteq \SSSi$. However, we prove $\cvx{\SSSb} \neq \RRRo$ and subsequently conjecture that $\cvx{\SSSb} = \cvx{\SSSi}$, based on numerical evidence.

For this \lcnamecref{sec:binary}, let $(\rv x, \rv y) \sim \DSBS{p}$ be a DSBS~\cite[Example~10.1]{ElGamal2011Network} with parameter $p \in [0,1]$, \ie, $\rv x \sim \bernoulli{\frac 12}$, $\rv x \perp \rv n \sim \bernoulli{p}$, and $\rv y \defas \rv x \oplus \rv n$. The cardinality bounds in \cref{pro:cardinality_bound} will enable us to use binary auxiliary random variables.

Subsequently we will provide evidence, supporting the following \lcnamecref{conj:loose_bound2}.
\begin{conjecture}
  \label{conj:loose_bound2}
  There exists $p \in [0,1]$, such that $\cvx{\SSSi} \neq \RRRo$.
\end{conjecture}

Let $\SSSb$ be defined as
\begin{align}
  \SSSb \defas \!\smash{\bigcup_{0 \le \alpha, \beta \le \frac 12}} \! \big\{
  (\mu, R_1, R_2):\,
  R_1 &\ge \log 2 - \binEnt{\alpha},\,  \eqnl
  R_2 &\ge \log 2 - \binEnt{\beta},\,  \eqnl
  \mu &\le \log 2 - \binEnt{\alpha * p * \beta} 
        \big\}.
        \label{eq:def-Sb}
\end{align}
By choosing $\rv u = \rv x \oplus \rv n_1$ and $\rv v = \rv y \oplus \rv n_2$, where $\rv n_1 \sim \bernoulli{\alpha}$ and $\rv n_2 \sim \bernoulli{\beta}$ are independent of $(\rv x, \rv y)$ and of each other,
it follows that $\SSSb \subseteq \SSSi$.
\begin{figure}[h]
  \centering
  \begin{tikzpicture}
    \begin{axis}[
      view={25}{30},
      grid=major,
      xlabel=$R_1$, ylabel=$R_2$, zlabel=$\mu$,
      extra x ticks={0.693147180559945}, extra x tick labels={$\log 2$},
      extra y ticks={0.693147180559945}, extra y tick labels={$\log 2$},
      extra z ticks={0.368064207168497}, extra z tick labels={$\mutInf{\rv x}{\rv y}$},
      zlabel style={rotate=-90,at={(axis description cs:-.05,.4)},anchor=east},
      ztick pos=right,
      xmin=0,
      xmax=0.693147180559945,
      ymin=0,
      ymax=0.693147180559945,
      zmin=-0.02,
      zmax=0.4,
      xtick={0.2,0.4},
      ytick={0.2,0.4},
      ztick={0,0.1,0.2},
      ]
      \addplot3[surf,
      mesh/ordering=y varies,
                        ]
      table {bin_region.dat};
    \end{axis}
  \end{tikzpicture}
  \caption{Boundary of $\SSSb$ for $p=0.1$.}
  \label{fig:Rprime}
\end{figure}
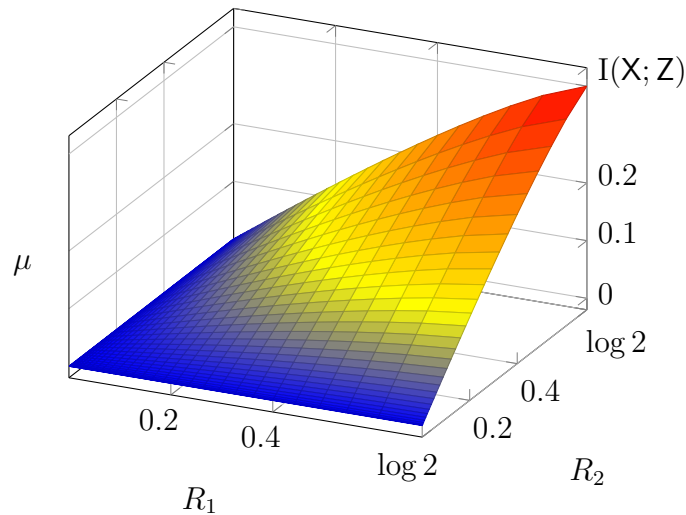
To illustrate the tradeoff between complexity ($R_1$, $R_2$) and relevance ($\mu$), the boundary of $\SSSb$ is depicted in \cref{fig:Rprime} for $p=0.1$.

Based on numerical experiments, we conjecture the following.
\begin{conjecture}
  \label{conj:binary}
  For the source $(\rv x, \rv y) \sim \DSBS{p}$ with $p \in [0,1]$ we have
  $\cvx{\SSSi} = \cvx{\SSSb}$, or equivalently $\SSSi \subseteq \cvx{\SSSb}$.
\end{conjecture}
The natural, stronger conjecture that $\SSSb = \SSSi$ already appeared in \cite[Conjecture~1,~Eq.~(14)]{Westover2008Achievable}.
However, there is the following counterexample \cite{Chapman2017}.
\begin{proposition}
  \label[pro]{pro:counterexample-conj1}
  For the source $(\rv x, \rv y) \sim \DSBS{0}$ we have $\SSSb \neq \SSSi$.
\end{proposition}
\begin{proof}
  \begin{figure}[h]
    \centering
    \begin{tikzpicture}
  [row sep=25mm,
  column sep=25mm,
  ampersand replacement=\&,
  line width=1pt,
  baseline=2mm,
  block/.style={draw, rectangle, minimum height=2em, rounded corners=2mm},
  dec/.style={draw, rectangle, minimum height=2em, rounded corners=2mm},
  symb/.style={outer sep=3mm},
  prob/.style={outer sep=-1mm},
  prob1/.style={outer sep=-.3mm},
  ]  \matrix{
    \coordinate  (u0); \&
    \coordinate  (x0); \&
    \coordinate  (v0); \\
    \coordinate  (u1); \&
    \coordinate  (x1); \&
    \coordinate  (v1); \\
  };
  \node [symb,anchor=south] at (u0) {$\rv u$};
  \node [symb,anchor=south] at (x0) {$\rv x = \rv y$};
  \node [symb,anchor=south] at (v0) {$\rv v$};
  \node [symb,anchor=east] at (u0) {$0$};
  \node [symb,anchor=east] at (u1) {$1$};
  \node [symb,anchor=west] at (v0) {$0$};
  \node [symb,anchor=west] at (v1) {$1$};

  \foreach \point in {u0,x0,v0,u1,x1,v1}
  \fill [black] (\point) circle (3pt);
  
  \foreach \pointa in {x0,x1}
  {
    \foreach \pointb in {u0,u1,v0,v1}
    \draw [-,postaction={decorate},decoration={
      markings,
      mark=at position 0.75 with {\arrow{latex}}}] (\pointa) to (\pointb);
  }

  \node [prob1,anchor=south west] at ($ (x0)!.75!(u0) $) {$a$};
  \node [prob1,anchor=south west] at ($ (x1)!.75!(u1) $) {$1$};
  \node [prob1,anchor=south east] at ($ (x0)!.75!(v0) $) {$0$};
  \node [prob1,anchor=south east] at ($ (x1)!.75!(v1) $) {$1-a$};

  \node [prob,anchor=south east] at ($ (x0)!.75!(u1) $) {$1-a$};
  \node [prob,anchor=south west] at ($ (x0)!.75!(v1) $) {$1$};
  \node [prob,anchor=north east] at ($ (x1)!.75!(u0) $) {$0$};
  \node [prob,anchor=north west] at ($ (x1)!.75!(v0) $) {$a$};

\end{tikzpicture}
    \caption{Random variables for the proof of \cref{pro:counterexample-conj1}.}
    \label{fig:bc-counterexample}
  \end{figure}
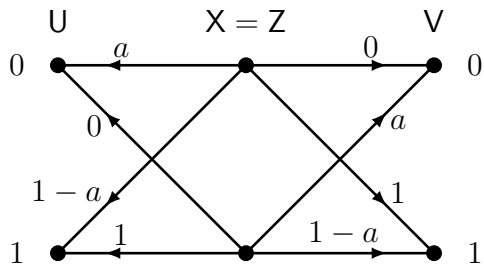
  For $a \in [0,1]$ we define $(\rv u, \rv v)$ by the binary channels depicted in \cref{fig:bc-counterexample}, satisfying ${\rv u \mkv \rv x \mkv \rv y \mkv \rv v}$.
  We obtain $(\mu, R, R) \in \SSSi$ with $R = \mutInf{\rv u}{\rv x} = \mutInf{\rv v}{\rv y} = \binEnt{\frac{a}{2}} - \frac{1}{2}\binEnt{a}$ and $\mu = \mutInf{\rv u}{\rv v} = 2R - a \log(2)$.
  \pgfkeys{/pgf/fixed point arithmetic}
  For $a=\pgfkeys{/pgf/number format/.cd,fixed,precision=5}\pgfmathprintnumber{\mya}$ we have $\mu \approx \pgfkeys{/pgf/number format/.cd,fixed,precision=6}\pgfmathprintnumber{\myIUV}$ and $R \approx \pgfkeys{/pgf/number format/.cd,fixed,precision=6}\pgfmathprintnumber{\myR}$. On the other hand, we obtain $\mu_{\mathrm{b}} \defas \max \{ \hat\mu : (\hat\mu, R, R) \in \SSSb \} < \pgfkeys{/pgf/number format/.cd,fixed,precision=6} \pgfmathparse{1e-6 * ceil(\myIUVbsc * 1e6)} \pgfmathprintnumber{\pgfmathresult}$ using \cref{eq:def-Sb} with $\alpha = \beta \approx \pgfkeys{/pgf/number format/.cd,fixed,precision=6}\pgfmathprintnumber{\myabsc}$. As $\mu_{\text{b}} < \mu$ we have $(\mu, R, R) \notin \SSSb$.
  
  This argument can be verified numerically using interval arithmetic \cite{Moore2009Introduction}. Code written in the Octave Programming Language \cite{GNUOctave} using its interval package \cite{OctaveInterval} can be found at \cite{Pichler2017DSBS}.
  \pgfkeys{/pgf/fpu=false}
\end{proof}

Note that \cref{pro:counterexample-conj1} does not impact \cref{conj:binary} as it concerns the case $p=0$. For $p=0$ we have $\rv x = \rv y$ and \cref{cor:comm-inf} implies $\ol\RRR = \big\{ (\mu, R_1, R_2) : R_1,R_2 \ge 0 \text{ and } \mu \le \min\{R_1,R_2,\log 2\} \big\}$. It is easily verified that $\ol\RRR = \cvx{\SSSb}$ and thus \cref{conj:binary} holds for $p=0$ by \cref{pro:cardinality_bound}.

In fact, it can be shown that the entire statement \cite[Conjecture~1]{Westover2008Achievable} does not hold.
The second part \cite[Conjecture~1,~Eq.~(15)]{Westover2008Achievable} claims that $\cvx{\SSSb} = \RRRo$. However, we shall construct a counterexample to this claim, showing that \cref{conj:loose_bound2} follows directly from \cref{conj:binary}.
\begin{proposition}
  \label[pro]{pro:binary-counterexample}
  For the source $(\rv x, \rv y) \sim \DSBS{\pgfkeys{/pgf/number format/.cd,fixed}\pgfmathprintnumber{\myp}}$, we have $\cvx{\SSSb} \neq \RRRo$.
\end{proposition}

To prove \cref{pro:binary-counterexample} we will construct a point $(\mu, R, R) \in \RRRo$ that satisfies $(\mu, R, R) \notin \cvx{\SSSb}$.
To this end, define the concave functions $\hat\mu_{\mathrm{b}}(R) \defas \max\{ \mu : (\mu, R, R) \in \cvx{\SSSb} \}$ and $\hat\mu_{\mathrm{o}}(R) \defas \max\{ \mu : (\mu, R, R) \in \RRRo \}$ for $R \in [0, \log 2]$.
In order to show $\cvx{\SSSb} \neq \RRRo$, it suffices to find $\hat R \in [0, \log 2]$ with $\hat\mu_{\mathrm{b}}(\hat R) < \hat\mu_{\mathrm{o}}(\hat R)$.

We can numerically compute an upper bound for the function $\hat\mu_{\mathrm{b}}$.
For $\alpha, \beta \in [0,\frac 12]$, we calculate
\begin{align}
  \wt R_1 &\defas \log 2 - \binEnt{\alpha} \label{eq:dsbs:wR1} , \\*
  \wt R_2 &\defas \log 2 - \binEnt{\beta} \label{eq:dsbs:wR2} , \text{ and} \\*
  \wt \mu &\defas \log 2 - \binEnt{\alpha * p * \beta}  \label{eq:dsbs:wmu}
\end{align}
on a suitably fine grid and upper bound the upper concave envelope of the implicitly defined function $\wt \mu(\wt R_1,\wt R_2)$. Evaluating it at $R=\wt R_1=\wt R_2$ yields an upper bound for $\hat\mu_{\mathrm{b}}(R)$.

On the other hand, we can obtain a lower bound for $\hat\mu_{\mathrm{o}}$ by computing \cref{eq:outer-bound:Ro} for specific pmfs that satisfy the Markov constraints in~\cref{thm:outer-bound}.
Note that based on the cardinality bound in~\cref{pro:cardinality_bound}, we can restrict the auxiliary random variables $\rv u$ and $\rv v$ to be binary.
We randomly sample the binary pmfs that satisfy the Markov constraints in~\cref{thm:outer-bound} (but not necessarily the long Markov chain $\rv u \mkv \rv x \mkv \rv y \mkv \rv v$) and in doing so encounter points strictly above the graph of $\hat\mu_{\mathrm{b}}$.
\begin{figure}[t]
  \centering
  \begin{tikzpicture}
    \begin{axis}[
            grid=major,
      xlabel=$R$, ylabel=$\mu$,
      xmin=0.673, xmax=0.694, ymin=0.358, ymax=0.369,
      try min ticks=4,
      max space between ticks=1000,
      extra x ticks={0.693147180559945}, extra x tick labels={$\log 2$},
      extra y ticks={0.368064207168497}, extra y tick labels={$\mutInf{\rv x}{\rv y}$},
      ylabel style={rotate=-90},
      legend pos=north west,
      ticklabel style={
                        /pgf/number format/precision=3,
                      },
      no markers,
      y label style={anchor=south},
      x label style={anchor=north},
      ]

      \addplot table
      {outer_bound_env.dat};
      \addlegendentry{$\hat\mu_{\mathrm{o}}$}

      \addplot table
      {inner_bound.dat};
      \addlegendentry{$\hat\mu_{\mathrm{b}}$}
    \end{axis}
  \end{tikzpicture}
  \pgfkeys{/pgf/number format/.cd,fixed}
  \caption{Numeric evaluation of $\hat\mu_{\mathrm{o}}$ and $\hat\mu_{\mathrm{b}}$ for $p=\pgfmathprintnumber{\myp}$.}
  \label{fig:binary}
\end{figure} %
\Cref{fig:binary} shows the resulting bounds for $p=\pgfkeys{/pgf/number format/.cd,fixed}\pgfmathprintnumber{\myp}$ in the vicinity of $R=\log 2$.
Albeit small, there is clearly a gap between $\hat\mu_{\mathrm{b}}$ and $\hat\mu_{\mathrm{o}}$ outside the margin of numerical error.
\begin{table}[ht]
  \pgfkeys{/pgf/number format/.cd,fixed,precision=100} %
  \resettabtoks %
  \foreach \u in {0,1} {
    \foreach \v in {0,1} { 
      \foreach \x in {0,1} {
        \foreach \z in {0,1} {
          \eaddtabtoks{ $\u$ }
          \addtabtoks{&}
          \eaddtabtoks{ $\v$ }
          \addtabtoks{&}
          \eaddtabtoks{ $\x$ }
          \addtabtoks{&}
          \eaddtabtoks{ $\z$ }
          \addtabtoks{&}
          \pgfmathparse{\Plgap[\u][\v][\x][\z]}
          \pgfmathroundto{\pgfmathresult}
          \eaddtabtoks{ $\pgfmathresult$ }
          \addtabtoks{\\} 
        }
      }
    }
  }%
  \centering
  \begin{tabular}[h]{cccc|l}
    \toprule $u$ & $v$ & $x$ & $y$ & $\Pcond{\rv u = u, \rv v = v}{\rv x = x, \rv y = y}$ \\
    \midrule
    \printtabtoks
    \bottomrule
  \end{tabular}
  \caption{Distribution resulting from random search.}
  \label{tab:dist_lgap}
\end{table}

\begin{proof*}[Proof of \cref{pro:binary-counterexample}]
  \pgfkeys{/pgf/number format/.cd,std,precision=6}%
    We observed the largest gap between the two bounds at a rate of $\hat R \approx \pgfmathprintnumber{\rateLGap}$. The particular distribution of $(\rv u, \rv v)$ at this rate, resulting from optimizing over the distributions that satisfy the Markov constraints in \cref{thm:outer-bound} is given in \cref{tab:dist_lgap} for reference. Note that this is an exact conditional pmf (\ie, not a numerical approximation) that satisfies the Markov chains $\rv u \mkv \rv x \mkv \rv y$ and $\rv x \mkv \rv y \mkv \rv v$.
  It achieves $\mutInf{\rv v}{\rv y} + \mutInf{\rv u}{\rv x} - \mutInf{\rv u \rv v}{\rv x \rv y} \approx \pgfmathprintnumber{\muLGap}$ which is $\Delta \approx \pgfmathprintnumber{\largestGap}$ above the inner bound, thus proving \cref{pro:binary-counterexample}.
  Using interval arithmetic \cite{Moore2009Introduction} this claim can be verified numerically. Code written in the Octave Programming Language \cite{GNUOctave} using its interval package \cite{OctaveInterval} can be found at \cite{Pichler2017DSBS}. It uses the distribution given in \cref{tab:dist_lgap}.  
\end{proof*}

We firmly believe that a tight characterization of the achievable region requires an improved outer bound. However, using current information theoretic tools, it appears very challenging to find a manageable outer bound based on the full Markov chain $\rv u \mkv \rv x \mkv \rv y \mkv \rv v$.

\begin{remark}
  Recently, Kumar and Courtade introduced a conjecture~\cite{Kumar2013Which,Courtade2014Which} concerning Boolean functions that maximize mutual information. Their work was inspired by a similar problem in computational biology~\cite{Klotz2014Canalizing}. A weaker form of their conjecture~\cite[Section~IV,~2)]{Courtade2014Which}, which was solved in~\cite{Pichler2016Tight}, corresponds to a zero-rate/one-bit variant of the binary example studied here.
\end{remark}

\section{The information bottleneck}
\label{sec:inform-bottl-1}

The information-theoretic problem posed by the IB method~\cite{Tishby2000Information} can be obtained as a special case from the biclustering problem. We will introduce the problem setup and subsequently show how it can be derived as a special case of \cref{def:biclustering:achievability}. Note that the definition slightly differs from~\cite[Definition~1]{Dhillon2003Information}. However, the achievable region is identical.

\begin{definition}
  \label[def]{def:ib:achievability}
  A pair $(\mu, R_1)$ is \emph{IB-achievable} if, for some $n \in \NN$, there exists $f\colon \XXX^n \to \MMM_1$ with $\log\card{\MMM_1} \le nR_1$ and
  \begin{align}
    \label{eq:ib:achievability}
    \mu \le \frac{1}{n} \mutInf{f(\rvt x)}{\rvt y} .
  \end{align}
  Let $\RRRib$ be the set of all IB-achievable pairs.
\end{definition}

\begin{proposition}
  \label[pro]{pro:equiv}
  For a pair $(\mu, R_1)$, the following are equivalent:
  \begin{enumerate}
  \item \label{itm:equiv:ib} $(\mu, R_1) \in \ol{\RRRib}$ .
    \item \label{itm:equiv:bi} $(\mu, R_1, \log\card{\YYY}) \in \ol{\RRR}$ .
  \item \label{itm:equiv:singleletter} There exists a random variable $\rv u$ such that $\rv u \mkv \rv x \mkv \rv y$, $\mutInf{\rv x}{\rv u} \le R$ and $\mutInf{\rv y}{\rv u} \ge \mu$.
  \end{enumerate}
\end{proposition}
\begin{proof}
  The equivalence `\labelcref{itm:equiv:ib} $\Leftrightarrow$ \labelcref{itm:equiv:bi}' holds as \cref{def:biclustering:achievability} collapses to \cref{def:ib:achievability} for $R_2 = \log\card{\YYY}$.
    To show `\labelcref{itm:equiv:bi} $\Leftrightarrow$ \labelcref{itm:equiv:singleletter}' apply \cref{thm:outer-bound,thm:inner-bound} with $\rv v = \rv y$.
\end{proof}

The tradeoff between `relevance' and `complexity' can equivalently be characterized by the IB function (\cf \cite{Courtade2014Multiterminal,Gilad2003Information}) $\mu_{\mathrm{IB}}(R) \defas \sup\{\mu : (\mu, R) \in \ol{\RRRib}\}$. \Cref{pro:equiv} provides
\begin{align}
    \mu_{\mathrm{IB}}(R) 
  = \max_{\substack{\rv u \;:\; \mutInf{\rv u}{\rv x} \le R \\
                                             \rv u \mkv \rv x \mkv \rv y}}
                       \mutInf{\rv u}{\rv y} . \label{eq:IB-function}
\end{align}
Interestingly, the function~\cref{eq:IB-function} is the solution to a variety of different problems in information theory.
As mentioned in~\cite{Gilad2003Information},~\cref{eq:IB-function} is the solution to the problem of loss-less source coding with one helper~\cite{Ahlswede1975Source,Wyner1975Source}.
Witsenhausen and Wyner~\cite{Witsenhausen1975Conditional} investigated a lower bound for a conditional entropy when simultaneously requiring another conditional entropy to fall below a threshold. Their work was a generalization of~\cite{Wyner1973theorema} and furthermore related 
to~\cite{Witsenhausen1974Entropy,Wyner1973theorem,Ahlswede1977connection,Ahlswede1975Source}.
The conditional entropy bound in~\cite{Witsenhausen1975Conditional} turns out to be an equivalent characterization of~\cref{eq:IB-function}.
Furthermore, $\mu_{\mathrm{IB}}$ characterizes the optimal error exponent, when testing against independence with one-sided data compression \cite[Theorem~2]{Ahlswede1986Hypothesis}.
Also in the context of gambling in the horse race market,~\cref{eq:IB-function} occurs as the maximum incremental growth in wealth when rate-limited side-information is available to the gambler~\cite[Theorem~3]{Erkip1998efficiency}.

\section{Multiple description CEO problem}
\label{sec:mult-descr-ceo}

In \cite[Appendix~B]{Courtade2014Multiterminal} Courtade and Weissman considered a multi-terminal extension of the IB problem, as introduced in \cref{sec:inform-bottl-1}, the CEO problem under an information constraint.
Analogous to how the IB problem is a special case of the biclustering problem (\cf \cref{pro:equiv}), this CEO problem presents a special case of a multi-terminal generalization of the biclustering problem~\cite{Pichler2016Distributed}.
Under a conditional independence assumption, Courtade and Weissman were able to provide a single letter characterization of the achievable region. In what follows we will extend this result, by incorporating MD coding for the CEO problem. Loosely speaking, we require the CEO to obtain valuable information from the message of just one agent alone. 
Surprisingly, this extension also permits a single-letter characterization under the same conditional independence assumption.

In what follows, let $(\rv x_\JJJ, \rv y)$ be $J+1$ random variables, where $\JJJ \defas \Ntoo{J}$, satisfying the Markov chain $\rv x_j \mkv \rv y \mkv \rv x_{\JJJ\setminus j}$ for every $j \in \JJJ$.
An $(n, R_\JJJ)$-code $f_\JJJ$ consists of $J$ functions $f_j\colon \XXX_j^n \to \MMM_j$ with $\log\card{\MMM_j} \le nR_j$ for every $j \in \JJJ$.

\begin{definition}
  \label[def]{def:ceo_achievable}
  A point $(\nu_0, \nu_\JJJ, R_\JJJ) \in \RR^{2J+1}$ is \emph{MI-achievable} if for some $n \in \NN$ 
  there exists an $(n, R_\JJJ)$-code $f_\JJJ$ such that 
  \begin{align}
    \label{eq:ceo_achievable}
    \frac{1}{n} \mutInf[big]{f_1(\rvt x_1), f_2(\rvt x_2), \dots, f_J(\rvt x_J)}{\rvt y} &\ge \nu_0 \text{ and } \\*
    \frac{1}{n} \mutInf[big]{f_j(\rvt x_j)}{\rvt y} &\ge \nu_j , & \shiftleft\text{ for all } j \in \JJJ  .
  \end{align}
  Denote the set of all MI-achievable points by $\RRRmi$.
\end{definition}

To shorten notation we will introduce the set of random variables
\begin{align}
  \PrIn \defas \left\{\rv u_\JJJ, \rv q :  \p[\rv q \rv u_\JJJ \rv x_\JJJ \rv y]{} = \p[\rv q]{} \cdot \p[\rv x_\JJJ \rv y]{} \cdot \prod_{j \in \JJJ} \pcond[\rv u_j][\rv x_j \rv q]{}{} \right\} .
\end{align}

\begin{definition}
  \label[def]{def:Ri}
  For a total order\footnote{For the notation regarding total orders refer to \cref{sec:notation}.} $\osub$ on $\JJJ$ and a set $\III \subseteq \JJJ$, let the region $\RRRmi^{(\osub, \III)} \subseteq \RR^{2J+1}$ be the set of tuples $(\nu_0, \nu_\JJJ, R_\JJJ)$ such that there exist random variables $(\rv u_\JJJ, \varnothing) \in \PrIn$ with
  \begin{align}
    R_j &\ge \condMutInf{\rv u_j}{\rv x_j}{\rv u_{\osup j}}, &&\shiftleft j \in \JJJ, \label{eq:md:rate} \\
    R_j &\ge \mutInf{\rv u_j}{\rv x_j}, &&\shiftleft j \in \III, \label{eq:md:rate2} \\
    \nu_{j} &\le \mutInf{\rv u_j}{\rv y}, &&\shiftleft j \in \JJJ, \label{eq:md:mu2} \\
    \nu_{j} &\le \condMutInf{\rv u_j}{\rv y}{\rv u_{\osup j}}, &&\shiftleft j \notin \III, \label{eq:md:mu} \\
    \nu_{0} &\le \mutInf{\rv u_\JJJ}{\rv y} . \label{eq:md:muJ}
    \end{align}
\end{definition}
\begin{remark}
  The purpose of the order $\osub$ is to determine the order of the messages for successive decoding. Equivalently, \cref{def:Ri} could be rephrased in terms of a permutation of $\JJJ$ in place of a total order.
\end{remark}

We are now able to state the single-letter characterization of $\ol{\RRRmi}$, the proof of which is provided in \cref{sec:proof:mdceo:main}.

\begin{theorem}
  \label{thm:mdceo:main}
  We have $\ol{\RRRmi} = \cvx{\bigcup_{\osub, \III} \RRRmi^{(\osub, \III)}}$, where the union is over all total orders   on $\JJJ$ and all sets $\III \subseteq \JJJ$.
\end{theorem}

  For $J=2$, we have $\ol{\RRRmi} = \cvx{\RRRmi^{(1)} \cup \RRRmi^{(2)} \cup \RRRmi^{(3)}}$, where $(\nu_0, \nu_\JJJ, R_\JJJ) \in \RRRmi^{(i)}$ iff, for some $(\rv u_\JJJ, \varnothing) \in \PrIn$, the following inequalities are satisfied
  \begin{align}
    \RRRmi^{(1)}:                           && \RRRmi^{(2)}:                            && \RRRmi^{(3)}: \nonumber\\
    \nu_1 &\le \mutInf{\rv y}{\rv u_1}                & \nu_1 &\le \condMutInf{\rv y}{\rv u_1}{\rv u_2}   & \nu_1 &\le \mutInf{\rv y}{\rv u_1} \\
    \nu_2 &\le \condMutInf{\rv y}{\rv u_2}{\rv u_1}   & \nu_2 &\le \mutInf{\rv y}{\rv u_2}                & \nu_2 &\le \mutInf{\rv y}{\rv u_2} \\
    \nu_0 &\le \mutInf{\rv y}{\rv u_1 \rv u_2}        & \nu_0 &\le \mutInf{\rv y}{\rv u_1 \rv u_2}        & \nu_0 &\le \mutInf{\rv y}{\rv u_1 \rv u_2} \\
    R_1 &\ge \mutInf{\rv u_1}{\rv x_1}                & R_1 &\ge \condMutInf{\rv u_1}{\rv x_1}{\rv u_2}   & R_1 &\ge \mutInf{\rv u_1}{\rv x_1} \\
    R_2 &\ge \condMutInf{\rv u_2}{\rv x_2}{\rv u_1}   & R_2 &\ge \mutInf{\rv u_2}{\rv x_2}                & R_2 &\ge \mutInf{\rv u_2}{\rv x_2} .
  \end{align}

  \begin{remark}
  Note that the total available rate of encoder $2$ is $R_2 = \condMutInf{\rv x_2}{\rv u_2}{\rv u_1}$ to achieve a point in $\RRRmi^{(1)}$. Interestingly, this rate is in general less than the rate required to ensure successful typicality decoding of $\rv u_2$. However, 
    $\nu_2 = \condMutInf{\rv y}{\rv u_2}{\rv u_1}$ can still be achieved.
\end{remark}
\begin{remark}
  \label{rmk:R_nu_dependence}
  On the other hand, fixing the random variables $\rv u_1$, $\rv u_2$ in the definition of $\RRRmi^{(i)}$ shows another interesting feature of this region. The achievable values for $\nu_1$ and $\nu_2$ vary across $i \in \{1,2,3\}$ and hence do not only depend on the chosen random variables $\rv u_1$ and $\rv u_2$, but also on the specific rates $R_1$ and $R_2$.  
\end{remark}
It is worth mentioning that by setting $\nu_j=0$ for $j=1,2,\dots,J$, the region $\ol{\RRRmi} $ reduces to the rate region in \cite[Appendix~B]{Courtade2014Multiterminal}.

The following \lcnamecref{pro:mdceo:cardinality} shows that $\RRRmi^{(\osub, \III)}$ is computable, at least in principle.
The given cardinality bound is not optimal, but it implies $\RRRmi^{(\osub, \III)} = \ol{\RRRmi^{(\osub, \III)}}$.
\begin{proposition}
  \label[pro]{pro:mdceo:cardinality}
  The region $\RRRmi^{(\osub, \III)}$ remains unchanged if the cardinality bound $\card{\UUU_j} \le \card{\rv x_j} + 4^J$ is imposed for every $j \in \JJJ$.
\end{proposition}
The proof of \cref{pro:mdceo:cardinality} is provided in \cref{sec:proof:mdceo:cardinality}.

\section{Summary and discussion}

We introduced a multi-terminal generalizations of the IB problem, termed information-theoretic biclustering. Interestingly, this problem is related to several other problems at the frontier of statistics and information theory and offers a formidable mathematical complexity.  Indeed, it is fundamentally different from `classical' distributed source coding problems where the encoders usually aim at reducing, as much as possible, redundant information among the sources while still satisfying a fidelity criterion. In the considered problem, however, the encoders are interested in maximizing precisely such redundant information. 

While an exact characterization of the achievable region is mathematically very challenging and still remains elusive, we provided outer and inner bounds to the set of achievable rates. We thoroughly studied the special case of two symmetric binary sources for which novel cardinality bounding techniques were developed. Based on numerical evidence we formulated a conjecture  that entails an explicit expression for the inner bound. This conjecture provides strong evidence  that our inner and outer bounds do not meet in general. We firmly believe that an improved outer bound, satisfying the adequate Markov chains, is required for a tight characterization of the achievable region.

Furthermore we considered an MD CEO problem which surprisingly permits a single-letter characterization of the achievable region. The resulting region has the remarkable feature that it allows to exploit rate that is in general insufficient to guarantee successful typicality encoding.

The interesting  challenge of the biclustering problem lies in the fact that one needs to bound the mutual information between two arbitrary encodings solely based on their rates. Standard information-theoretic manipulations seem incapable of handling this requirement well.

\section*{Funding}
Wiener Wissenschafts-, Forschungs- und Technologiefonds (ICT12-054, ICT15-119);
This project has received funding from the European Union’s Horizon 2020 research and innovation programme under the Marie Skłodowska-Curie grant agreement No 792464.

\section*{Acknowledgment}

The authors would like to thank Shlomo Shamai (Shitz) and Emre Telatar for insightful discussions regarding the binary example.
We would also like to thank Christian Chapman from the School of Electrical, Computer and Energy Engineering at the Arizona State University for providing the counterexample used in the proof of \cref{pro:counterexample-conj1}.

\section*{Data availability statement}
No new data were generated or analysed in support of this review.

\appendix

\section{Proofs}
\label{apx:proofs}

\subsection{Proof of \texorpdfstring{\cref{thm:equiv}}{Theorem \ref{thm:equiv}}}
\label{apx:proof:equiv}

To prove $\RRR \subseteq \RRR_*$, assume $(\mu, R_1, R_2) \in \RRR$ and choose $n$, $f$ and $g$ according to \cref{def:biclustering:achievability}. Defining $\rv u \defas f(\rvt x)$ and $\rv v \defas g(\rvt y)$ yields inequalities \cref{eq:multi-letter:1,eq:multi-letter:2,eq:multi-letter:mu} and satisfies the required Markov chain.

The inclusions $\ol{\RRR_*} \subseteq \ol{\RRRht}$ and $\ol{\RRR_*} \subseteq \ol{\RRRpr}$ follow by applying the achievability results \cite[Corollary~6]{Han1987Hypothesis} and \cite[Theorem~1]{Westover2008Achievable}, respectively, to the vector source $(\rvt x, \rvt y)$.

Assuming $(\mu, R_1, R_2) \in \RRRht$, choose an arbitrary $\eps > 0$ and pick an $(n,R_1,R_2)$ hypothesis test $(f_n,g_n,\AAA_n)$ such that $\alpha_{n} \le \eps$ and $-\log\beta_{n} \ge n(\mu-\eps)$.
We apply the log-sum inequality \cite[Theorem~2.7.1]{Cover2006Elements} and obtain for any $\eps\prm > 0$, provided that $\eps$ is small enough and $n$ is large enough,%
\begin{align}
  \mutInf[big]{f(\rvt x)}{g(\rvt y)} &\ge (1-\alpha_n) \log\frac{1-\alpha_n}{\beta_n} + \alpha_n \log\frac{\alpha_n}{1-\beta_n} 
        \ge n(\mu - \eps\prm) , \label{eq:hypo:proof}
\end{align}
which implies $(\mu, R_1, R_2) \in \ol{\RRR}$.

Similarly, assume $(\mu, R_1, R_2) \in \RRRpr$ and for an arbitrary $\eps > 0$ and sufficiently large $n \in \NN$ pick an $(n,R_1,R_2)$-code $(f,g)$ and $\phi$ satisfying \cref{eq:pattern_condition}. Then (using the notation of \cref{def:pr}),
\begin{align}
  \mutInf[big]{f(\rvt x)}{g(\rvt y)} &= \condMutInf[big]{\CB}{g(\rvt y(\rv w))}{\rv w} \label{eq:pattern:1}\\
                                         &\ge \condMutInf[big]{\CB}{\rv w}{g(\rvt y(\rv w))} \label{eq:pattern:2} \\
                                                                          &\ge n\mu - \condEnt[big]{\rv w}{\phi\big(\CB, g(\rvt y(\rv w))\big)}   \label{eq:pattern:3}\\
                                     &\stacklap{\cref{eq:pattern_condition}}{\ge} n\mu - \binEnt{\eps} - \eps n \mu .  \label{eq:fano}
\end{align}
The equality in \cref{eq:pattern:1} holds as $\rvt x(i) \perp \rvt y(j)$ for $i \neq j$, \cref{eq:pattern:2} follows from $\rv w \perp \CB$, and \cref{eq:pattern:3} follows from $\rv w \perp \rvt y(\rv w)$, the fact that $\ent{\rv w} = n\mu$, and the data processing inequality~\cite[Theorem~2.8.1]{Cover2006Elements}. Fano's inequality~\cite[Theorem~2.10.1]{Cover2006Elements} was used in~\cref{eq:fano}. This shows $(\mu, R_1, R_2) \in \ol{\RRR}$ as $\eps$ was arbitrary.

\subsection{Proof of \texorpdfstring{\cref{pro:cardinality_bound}}{Proposition \ref{pro:cardinality_bound}}}
\label{apx:proof:cardinality_bound}

We start with the proof of $\cvx{\SSSo} = \RRRo$.
 
For fixed random variables $(\rv x, \rv y)$ define the set of pmfs (with finite, but arbitrarily large support)
\begin{align}
  \QQQ \defas \{\p[\rv u, \rv x, \rv y, \rv v]{} : \rv u \mkv \rv x \mkv \rv y, \rv x \mkv \rv y \mkv \rv v\} ,
\end{align}
and the compact set of pmfs with fixed alphabet size
\begin{align}
  \QQQ(a,b) \defas \{\p[\rv u, \rv x, \rv y, \rv v]{} \in \QQQ : \card{\UUU} = a, \card{\VVV} = b \} .
\end{align}
Define the continuous vector valued function $\vt F \defas (F_1, F_2, F_3)$ as
\begin{align}
  F_1(\p[\rv u, \rv x, \rv y, \rv v]{}) &\defas
                                        \mutInf{\rv x}{\rv u} + \mutInf{\rv y}{\rv v} - \mutInf{\rv u\rv v}{\rv x \rv y} ,
\\
  F_2(\p[\rv u, \rv x, \rv y, \rv v]{}) &\defas \mutInf{\rv u}{\rv x}, \\
  F_3(\p[\rv u, \rv x, \rv y, \rv v]{}) &\defas \mutInf{\rv v}{\rv y} .
\end{align}

We can now write 
$\RRRo = \vt F(\QQQ) + \OOO$ and $\SSSo = \vt F\big(\QQQ(\card[normal]{\XXX},\card[normal]{\YYY})\big) + \OOO$ where $\OOO \defas (\RR_- \times \RR_+ \times \RR_+)$.
Since $\RRRo$ is convex, we may define the extended real function $\psi(\vt\lambda) \defas \inf_{\vt x \in \RRRo} \vt\lambda \cdot \vt x$ and obtain \cite[Theorem~2.2,~3.]{Gruenbaum2003Convex}
\begin{align}
  \ol{\cvx{\RRRo}} = \ol{\RRRo} = \bigcap_{\vt\lambda \in \RR^3} \Big\{ \vt x \in \RR^3\!: \vt x \cdot \vt\lambda \ge \psi(\vt\lambda) \Big\} .
\end{align}
From the definition of $\RRRo$, we have $\psi(\vt \lambda) = -\infty$ if $\vt\lambda \notin \OOO$, and $\psi( \vt\lambda ) = \inf_{\p{} \in \QQQ} \vt\lambda \cdot \vt F(\p{})$ otherwise. This shows, that
\begin{align}
  \ol{\RRRo} = \bigcap_{\vt\lambda \in \OOO} \Big\{ \vt x \in \RR^3 : \vt x \cdot \vt\lambda \ge \psi(\vt\lambda) \Big\} \label{eq:bincard:Ro_characterization}
\end{align}
and using the same argument, one can also show that
\begin{align}
  \ol{\cvx{\SSSo}} &= \bigcap_{\vt\lambda \in \OOO} \Big\{ \vt x \in \RR^3 : \vt x \cdot \vt\lambda \ge \wt\psi(\vt\lambda) \Big\} , & \wt\psi(\vt\lambda) &= \min_{\p{} \in \QQQ(\card{\XXX}, \card{\YYY})} \vt\lambda \cdot \vt F(\p{}) .\label{eq:bincard:So_characterization}
\end{align}
We shall now prove that $\psi(\vt\lambda) = \wt\psi(\vt\lambda)$ for $\vt\lambda \in \OOO$.
For arbitrary $\vt\lambda \in \OOO$ and $\delta > 0$, we can find random variables $(\wrv u, \rv x, \rv y, \wrv v) \sim \wtp{} \in \QQQ$ with $\vt\lambda \cdot \vt F(\wtp{}) \le \psi(\vt\lambda) + \delta$. 
By compactness of $\QQQ(a,b)$ and continuity of $\vt F$, there is
$\p{} \in \QQQ(\card[normal]{\wt\UUU},\card[normal]{\wt\VVV})$ with 
\begin{align}
  \vt\lambda \cdot \vt F(\p{}) &= \min_{\htp{} \in \QQQ(\card[normal]{\wt\UUU},\card[normal]{\wt\VVV})} \vt\lambda \cdot \vt F(\htp{})
    \le \vt\lambda \cdot \vt F(\wtp{}) 
    \le \psi(\vt\lambda) + \delta . \label{eq:bincard:max_p}
\end{align}
We now show that there exists $\htp{} \in \QQQ(\card{\XXX}, \card{\YYY})$ with 
\begin{align}
  \vt\lambda \cdot \vt F(\htp{}) = \vt\lambda \cdot \vt F(\p{}) . \label{eq:bincard:complete}
\end{align}
As a consequence of the inequalities $F_1 \le F_2$ and $F_1 \le F_3$ we have $\vt\lambda \cdot \vt F(\p{}) = 0$ if $\lambda_1 + \max\{\lambda_2, \lambda_3\} \ge 0$.
Thus, we only need to show~\cref{eq:bincard:complete} for $\vt\lambda \in \OOO$ with $\lambda_1 + \lambda_2 < 0$ and $\lambda_1 + \lambda_3 < 0$. To this end we use the perturbation method~\cite{Gohari2012Evaluation,Jog2010information} and perturb $\p{}$, obtaining the candidate
\begin{align}
  (\hrv u, \rv x, \rv y, \hrv v) \sim  \htp{u,x,z,v} = \p{u,x,z,v}\big(1+\eps\phi(u)\big) .
\end{align}
We require
\begin{align}
  1+\eps\phi(u) &\ge 0 , &&\shiftleft \text{ for every } u \in \UUU ,\label{eq:bincard:phi_rq1} \\ 
  \Exp{\phi(\rv u)} &= 0 , \label{eq:bincard:phi_rq2} \\
  \condExp{\phi(\rv u)}{\rv x=x, \rv y=z} &= 0 , &&\shiftleft \text{ if } \p{x,z} > 0 . \label{eq:bincard:phi_rq3}
\end{align}
The conditions~\cref{eq:bincard:phi_rq1,eq:bincard:phi_rq2} ensure that $\htp{}$ is a valid pmf and~\cref{eq:bincard:phi_rq3} implies $\htp{} \in \QQQ$.
Observe that there is an $\eps_0 > 0$ for any $\phi$, such that~\cref{eq:bincard:phi_rq1} is satisfied for $\eps \in [-\eps_0, \eps_0]$. Furthermore,~\cref{eq:bincard:phi_rq3} is equivalent to
\begin{align}
  \condExp{\phi(\rv u)}{\rv x = x} &= 0, \quad \text{ for every } x \in \XXX \label{eq:bincard:phi_rq4}
\end{align}
due to the Markov chain $\rv u \mkv \rv x \mkv \rv y$. Note also that \cref{eq:bincard:phi_rq4} already implies \cref{eq:bincard:phi_rq2}. If $\card{\UUU} \ge \card{\XXX} + 1$ there is a non-trivial solution to~\cref{eq:bincard:phi_rq4}, which means there exists $\phi \not\equiv 0$ such that~\cref{eq:bincard:phi_rq1,eq:bincard:phi_rq2,eq:bincard:phi_rq3} are satisfied.
We have
\begin{align}
  \vt\lambda \cdot \vt F(\htp{}) &= \lambda_1 \Big[ \mutInf{\rv x}{\rv u} - \mutInf{\rv u \rv v}{\rv x \rv y} + \ent{\rv y}
                                   + \eps \entPhi{\rv u} - \eps \entPhi{\rv u \rv x}
                                   \eqnl&\qquad- \eps \entPhi{\rv u \rv v} + \eps \entPhi{\rv u \rv x \rv y \rv v}
                                          + \ent[normal]{\hrv v} - \ent[normal]{\rv y \hrv v} \Big]
                                          \eqnl&\quad + \lambda_2 [ \mutInf{\rv x}{\rv u} + \eps \entPhi{\rv u} - \eps \entPhi{\rv u \rv x} ]
                                                 \eqnl&\quad + \lambda_3 [ \ent{\rv y} + \ent[normal]{\hrv v} - \ent[normal]{\rv y \hrv v} ] . \label{eq:bincard:2}
\end{align}
Here, we used the shorthand $\entPhi{\rv u \rv x} \defas - \sum_{u,x} \p{u,x} \phi(u) \log\p{u,x}$ and analogous for other combinations of random variables. 
By~\cref{eq:bincard:max_p}, we have $\frac{\partial^2}{\partial\eps^2} \vt\lambda \cdot \vt F(\htp{})\big|_{\eps = 0} \ge 0$.
Observe that
\begin{align}
  &\frac{\partial}{\partial\eps} \big( \ent[normal]{\hrv v} - \ent[normal]{\rv y \hrv v} \big)  
    = \frac{\partial}{\partial\eps} \sum_{z,v} \htp{z,v} \log{\frac{\htp{z,v}}{\htp{v}}} \\
          &\qquad= \sum_{z,v} \frac{\partial \htp{z,v}}{\partial\eps} \log{\frac{\htp{z,v}}{\htp{v}}} + 
    \frac{\partial \htp{z,v}}{\partial\eps} - \frac{\htp{z,v} \frac{\partial \htp{v}}{\partial\eps}}{\htp{v}}
\end{align}
and consequently,
\begin{align}
  & \frac{\partial^2}{\partial\eps^2} \vt\lambda \cdot \vt F(\htp{}) = (\lambda_1 + \lambda_3) \frac{\partial^2}{\partial\eps^2} (\ent[normal]{\hrv v} - \ent[normal]{\rv y \hrv v}) \\
        &\qquad= (\lambda_1 + \lambda_3) \sum_{z,v} \left(\frac{\partial \htp{z,v}}{\partial\eps}\right)^2 \frac{1}{\htp{z,v}} 
    \nonumber\\&\qquad\qquad - 2\frac{\partial \htp{z,v}}{\partial\eps} \frac{\partial \htp{v}}{\partial\eps} \frac{1}{\htp{v}}
  + \left(\frac{\partial \htp{v}}{\partial\eps}\right)^2  \frac{\htp{z,v}}{\htp{v}^2} .
\end{align}
Here we already used that $\frac{\partial^2 \htp{v}}{\partial\eps^2} \equiv \frac{\partial^2 \htp{z,v}}{\partial\eps^2} \equiv 0$. It is straightforward to calculate
\begin{align}
  \frac{\partial \htp{v}}{\partial\eps} &= \p{v}\condExp{\phi(\rv u)}{\rv v=v}, \\
  \frac{\partial \htp{z,v}}{\partial\eps} &= \p{z,v}\condExp{\phi(\rv u)}{\rv v=v,\rv y=z}, \\
  \htp{z,v}|_{\eps=0} &= \p{z,v}, \\
  \htp{v}|_{\eps=0} &= \p{v},
\end{align}
and thus, taking into account that $\lambda_1 + \lambda_3 < 0$,
\begin{align}
  0 &\ge \sum_{z,v} \p{z,v} \big(\condExp{\phi(\rv u)}{\rv v=v,\rv y=z} - \condExp{\phi(\rv u)}{\rv v=v}\big)^2 ,
\end{align}
which implies for any $(z,v) \in \YYY \times \VVV$ with $\p{z,v} > 0$,
\begin{align}
  \sum_u \pcond{u}{z,v} \phi(u) = \sum_u \pcond{u}{v} \phi(u) . \label{eq:bincard:3}
\end{align}
From~\cref{eq:bincard:3} we can conclude
\begin{align}
  &\ent[normal]{\hrv v} - \ent[normal]{\rv y \hrv v} = \sum_{z,v} \htp{z,v} \log\frac{\htp{z,v}}{\htp{v}} \\
  &\qquad\RA= \sum_{z,v,u} \p{u,z,v}(1+\eps\phi(u)) \log\frac{ \sum_{\hat{u}} \p{\hat{u},z,v} \big(1+\eps\phi(\hat{u})\big) }{ \sum_{\hat{u}} \p{\hat{u},v} \big(1+\eps\phi(\hat{u})\big)} \\
    &\qquad\RA= \sum_{z,v,u} \p{u,z,v}(1+\eps\phi(u)) \log\frac{ \p{z,v} \big(1 + \eps \sum_{\hat{u}} \pcond{\hat{u}}{z,v} \phi(\hat{u})\big) }{ \p{v} \big(1+ \eps \sum_{\hat{u}} \pcond{\hat{u}}{v} \phi(\hat{u})\big)} \\
    &\qquad\RA[\cref{eq:bincard:3}]= \sum_{z,v,u} \p{u,z,v}(1+\eps\phi(u)) \log\frac{\p{z,v}}{\p{v}} \\
  &\qquad\RA= \sum_{z,v} \p{z,v} \log\frac{\p{z,v}}{\p{v}} + \eps \sum_{z,v,u} \phi(u) \p{u,z,v} \log\frac{\p{z,v}}{\p{v}} \\
  &\qquad\RA= \ent{\rv v} - \ent{\rv y \rv v} + \eps \entPhi{\rv v} - \eps \entPhi{\rv y \rv v} ,
\end{align}
where we used 
\begin{align}
  \entPhi{\rv v} &\defas - \sum_{u,v} \p{u,v} \phi(u) \log\p{v} , \\
  \entPhi{\rv y \rv v} &\defas - \sum_{u,z,v} \p{u,z,v} \phi(u) \log\p{z,v} .
\end{align}
Substituting in~\cref{eq:bincard:2} shows that $\vt\lambda \cdot \vt F(\htp{})$ is linear in $\eps$. And by the optimality of $\p{}$ it must be constant. We may now choose $\eps$ maximal, \ie, such that there is at least one $u \in \UUU$ with $\p{u} (1 + \eps\phi(u)) = 0$. This effectively reduces the cardinality of $\hat{\UUU}$ by at least one and may be repeated until $\card[normal]{\hat{\UUU}} = \card{\XXX}$ (as then $\phi \equiv 0$). The same process can be carried out for $\rv v$ and yields $\htp{} \in \QQQ(\card{\XXX}, \card{\YYY})$ such that~\cref{eq:bincard:complete} holds.

Using \cref{eq:bincard:complete,eq:bincard:max_p} we obtain
\begin{align}
  \psi(\vt\lambda) \le \wt\psi(\vt\lambda) \le \vt\lambda \cdot \vt F(\htp{})
    \le \psi(\vt\lambda) + \delta .
\end{align}
As $\delta > 0$ was arbitrary,
we proved $\psi(\vt\lambda) = \wt\psi(\vt\lambda)$, which implies $\ol{\RRRo} = \ol{\cvx{\SSSo}}$ using \cref{eq:bincard:Ro_characterization,eq:bincard:So_characterization}.
We find $\RRRo = \cvx{\SSSo}$ by writing
\begin{align}
  \ol{\RRRo} &= \ol{\cvx{\SSSo}} = \ol{\cvx{\vt F(\QQQ(\card{\XXX}, \card{\YYY})) + \OOO}} \\
                          &= \cvx{\SSSo} = \cvx{\vt F(\QQQ(\card{\XXX}, \card{\YYY}))} + \OOO  \label{eq:bincard:apply_conv_sum}\\
             &\subseteq \vt F(\QQQ) + \OOO \label{eq:bincard:FQ}\\
             &= \RRRo ,
\end{align}
where \cref{eq:bincard:apply_conv_sum} follows from \cref{lem:convex_sum}. The relation \cref{eq:bincard:FQ} is a consequence of $\QQQ(\card{\XXX}, \card{\YYY}) \subseteq \QQQ$ and the convexity of $\vt F(\QQQ)$.

In order to prove $\cvx{\SSSi}=\cvx{\RRRi}$, we will only show the cardinality bound $\card{\UUU} \le \card{\XXX}$. The corresponding bound for $\card{\VVV}$ follows analogously.
We note that the weaker bounds $\card{\UUU} \le \card{\XXX} + 1$ and $\card{\VVV} \le \card{\YYY} + 1$
can be obtained directly using the convex cover method \cite[Appendix~C]{ElGamal2011Network}, \cite{Ahlswede1975Source,Wyner1976rate}.
Define the continuous vector-valued function 
\begin{align}
  \vt F(\p[\wrv u, \wrv x, \wrv y, \wrv v]{}) \defas \big(\mutInf[normal]{\wrv u}{\wrv v}, \mutInf[normal]{\wrv x}{\wrv u}, \mutInf[normal]{\wrv y}{\wrv v}\big) ,  
\end{align}
and the compact, connected sets of pmfs 
\begin{align}
  \QQQ &\defas \Big\{\p[\wrv u, \wrv x, \wrv y, \wrv v]{} : \p[\wrv u, \wrv x, \wrv y, \wrv v]{} = \pcond[\wrv u][\rv x]{}{}\p[\rv x, \rv y]{}\pcond[\wrv v][\rv y]{}{}, \wt\UUU = \NZtoo[big]{\card{\XXX}}, \wt\VVV = \NZtoo[big]{\card{\YYY}} \Big\} ,\\
  \wt\QQQ &\defas \Big\{\p[\wrv u, \wrv x, \wrv y, \wrv v]{} \in \QQQ : \wt\UUU = \Ntoo[big]{\card[normal]{\XXX}} \Big\} .
\end{align}
To complete the proof of the~\lcnamecref{pro:cardinality_bound}, it suffices to show 
\begin{align}
  \label{eq:proof:cardinality_bound:cvx_subset}
  \cvx[big]{\vt F(\QQQ)} \subseteq \cvx[big]{\vt F(\wt\QQQ)} ,
\end{align}
since we then have with $\OOO \defas (\RR_- \times \RR_+ \times \RR_+)$,
\begin{align}
  \cvx{\RRRi} &= \cvx[big]{ \vt F(\QQQ) + \OOO} \\
              &= \cvx[big]{\vt F(\QQQ)} + \OOO \label{eq:proof:cardinality_bound:conv_sum} \\
              &\stackrel{\mathclap{\cref{eq:proof:cardinality_bound:cvx_subset}}}{\subseteq} \cvx[big]{ \vt F(\wt\QQQ)} + \OOO \\
              &= \cvx[big]{\vt F(\wt\QQQ) + \OOO} \label{eq:proof:cardinality_bound:conv_sum2}\\
              &= \cvx{\SSSi} ,
\end{align}
where \cref{eq:proof:cardinality_bound:conv_sum,eq:proof:cardinality_bound:conv_sum2} follow from~\cref{lem:convex_sum}.
The region $\vt F(\QQQ) \subseteq \RR^3$ is compact~\cite[Theorem~4.22]{Rudin1976Principles}.
Therefore, its convex hull $\cvx[big]{\vt F(\QQQ)}$ is compact~\cite[Corollary~5.33]{Aliprantis2006Infinite} and can be represented as an intersection of halfspaces in the following manner~\cite[Proposition~2.2,~3.]{Gruenbaum2003Convex}:
Defining $V(\vt\lambda) \defas \max_{\vt x \in \vt F(\QQQ)} \vt\lambda \cdot \vt x$ for $\vt\lambda = (\lambda_1, \lambda_2, \lambda_3) \in \RR^3$, we have 
\begin{align}
  \label{eq:proof:cardinality_bound:cvx_P}
  \cvx[big]{\vt F(\QQQ)} = \bigcap_{\vt\lambda \in \RR^3} \Big\{\vt x \in \RR^3 : \vt\lambda \cdot \vt x \le V(\vt\lambda)\Big\} .  
\end{align}
With the same reasoning we obtain 
\begin{align}
  \label{eq:proof:cardinality_bound:cvx_Pprime}
  \cvx[big]{\vt F(\wt\QQQ)} = \bigcap_{\vt\lambda \in \RR^3} \Big\{\vt x \in \RR^3 : \vt\lambda \cdot \vt x \le \wt V(\vt\lambda)\Big\} ,
\end{align}
where $\wt V(\vt\lambda) \defas \max_{\vt x \in \vt F(\wt\QQQ)} \vt\lambda \cdot \vt x$.
We next show $\wt V(\vt\lambda) \ge V(\vt\lambda)$ which already implies~\cref{eq:proof:cardinality_bound:cvx_subset} due to~\cref{eq:proof:cardinality_bound:cvx_P,eq:proof:cardinality_bound:cvx_Pprime}.

Let $\vt t = (t_x)_{x \in \XXX \setminus x_0}$ be $\card{\XXX} - 1$ test functions $t_x(\p[\wrv x]{}) \defas \p[\wrv x]{x}$ for all but one $x \in \XXX$. Choose any $\vt\lambda \in \RR^3$ and fix $(\rv u, \rv x, \rv y, \rv v) \sim \p{} \in \QQQ$ that achieve $\vt\lambda \cdot \vt F(\p{}) = V(\vt\lambda)$. Define the continuous function 
\begin{align}
  f(\p[\wrv x]{}) &\defas \lambda_1(\ent{\rv v} - \ent[normal]{\wrv v}) + \lambda_2(\ent{\rv x} - \ent[normal]{\wrv x}) + \lambda_3 \mutInf{\rv y}{\rv v} 
\end{align}
where $(\wrv v, \wrv y, \wrv x) \sim \pcond[\rv v][\rv y]{}{}\pcond[\rv y][\rv x]{}{}\p[\wrv x]{}$.
The point $\big( (\p[\rv x]{x})_{x \in \XXX \setminus x_0}, V(\vt\lambda) \big)$ lies in the convex hull of the compact~\cite[Theorem~26.5]{Munkres2000Topology} and connected~\cite[Theorem~4.22]{Rudin1976Principles} set $(\vt t, f)\big(\QQQ\big)$. Therefore, by the strengthened Carathéodory theorem~\cite[Theorem~18(ii)]{Eggleston1958Convexity}, $\card{\XXX}$ points suffice, \ie, there exists a random variable $\rv u\prm$ with $\card{\UUU\prm} = \card{\XXX}$ and thus $\p[\rv u\prm, \rv x, \rv y, \rv v]{} \in \wt\QQQ$, such that $\Exp[\rv u\prm]{f(\pcond[\rv x][\rv u\prm]{\wc}{\rv u\prm})} = \vt\lambda \cdot \vt F(\p[\rv u\prm, \rv x, \rv y, \rv v]{}) = V(\vt\lambda)$. This shows $\wt V(\vt\lambda) \ge V(\vt\lambda)$.

By applying the same reasoning to $\rv v$, one can show that $\card{\VVV} = \card{\YYY}$ also suffices.

\subsection{Proof of \texorpdfstring{\cref{thm:mdceo:main}}{Theorem \ref{thm:mdceo:main}}}
\label{sec:proof:mdceo:main}

We will prove \cref{thm:mdceo:main}, by showing an inner and an outer bound (\cref{lemma:mdceo:achiev,lem:mdceo:converse}, respectively) and subsequently prove tightness.
\begin{lemma}
  \label{lemma:mdceo:achiev}
  We have $\RRRmi^{(\osub, \III)} \subseteq \ol{\RRRmi}$ for any $\III \subseteq \JJJ$ and any total order $\osub$ on $\JJJ$.
\end{lemma}
\begin{proof}
  In part, the proof of this \lcnamecref{lemma:mdceo:achiev} closely follows the proof of~\cite[Theorem~1]{Han1980Unified}. We will use $\typ[n][\eps]{\rv x}$ to denote the $\eps$ typical sequences~\cite[Section~III]{Han1980Unified}.
  
  Pick a total order $\osub$ on $\JJJ$, a set $\III \subseteq \JJJ$, $(\rv u_\JJJ, \varnothing) \in \PrIn$, and $(\nu_0, \nu_\JJJ, R_\JJJ)$ satisfying \cref{eq:md:rate,eq:md:rate2,eq:md:mu,eq:md:mu2,eq:md:muJ}. We will use typicality coding and deterministic binning to obtain a code. Letting $\hat R_j=\condMutInf{\rv u_j}{\rv x_j}{\rv u_{\osup j}}$, we verify for $\AAA = \osupeq \jzidx$, $\jzidx \in \JJJ$, and any $\BBB \subseteq \AAA$, that
\begin{align}
  \sum_{j \in \BBB} \hat R_j &= \sum_{j \in \BBB} \condMutInf{\rv u_j}{\rv x_j}{\rv u_{\osup j}} \\
                             &\ge \sum_{j \in \BBB} \condMutInf{\rv u_j}{\rv x_{\BBB}}{\rv u_{\osup j}, \rv u_{\AAA\setminus \BBB}} \\
                             &= \condMutInf{\rv u_{\BBB}}{\rv x_{\BBB}}{\rv u_{\AAA\setminus \BBB}} . \label{eq:b_tung:constraint}
\end{align}
Following the proof of~\cite[Theorem~1]{Han1980Unified} and applying the conditional typicality lemma~\cite[Lemma~3.1.(iv)]{Han1980Unified}, we can thus for any $\eps > 0$ and $n$ large enough obtain an $(n, \hat R_\JJJ + \eps)$-code $\hat f_\JJJ$ and for any $\AAA = \osupeq j$, $j \in \JJJ$, a decoding function $g_\AAA$, such that, $\Prob{\SSS_\AAA} \ge 1-\eps$, where $\SSS_\AAA$ is the `success' event  
\begin{align}
  \SSS_\AAA \defas \{(\rvt y, \rvt x_\JJJ, g_\AAA \circ \hat f_\AAA(\rvt x_\AAA)) \in \typ[n][\eps]{\rv y, \rv x_\JJJ, \rv u_\AAA}\}.
\end{align}
For $j \notin \III$, we set $f_j \defas \hat f_j$, but for $j \in \III$, we let $f_j$ be typicality encoding \emph{without} binning, in total yielding an $(n, R_\JJJ + \eps)$-code. Moreover, for $n$ large enough and $j \in \III$, we find decoding functions $g_j$, such that $\Prob{\SSS_j} \ge 1-\eps$ also for the `success' events
\begin{align}
  \SSS_j \defas \{ (\rvt y, \rvt x_\JJJ, g_j \circ f_j(\rvt x_j)) \notin \typ[n][\eps]{\rv y, \rv x_\JJJ, \rv u_j} \} .
\end{align}

To shorten notation, let $\rv w_j = f_j(\rvt x_j)$ and $\hrv w_j \defas \hat f_j(\rvt x_j)$ for $j \in \JJJ$.
Pick an arbitrary\footnote{In what follows, we will routinely merge expressions that can be made arbitrarily small (for $n$ large and $\eps$ sufficiently small) and bound them by $\eps\prm$.} $\eps\prm > 0$. Provided that $n$ is large enough and $\eps$ small enough, we have for any $\AAA = \osupeq j$ ($j \in \JJJ$),
\begin{align}
  \frac 1n \mutInf{\rvt y}{\rv w_\AAA} &\ge \frac 1n \mutInf[big]{\rvt y}{\hrv w_\AAA} \label{eq:data_proc1} \\
                                       &\ge \frac 1n \mutInf[big]{\rvt y}{g_\AAA(\hrv w_\AAA)} \label{eq:data_proc2} \\
                                       &= \ent{\rv y} - \frac 1n \condEnt[big]{\rvt y}{g_\AAA(\hrv w_\AAA)}\\
                                       & \ge \ent{\rv y} - \frac 1n \condEnt[big]{\rvt y, \ind{\SSS_\AAA}{}}{g_\AAA(\hrv w_\AAA)}\\
                                       &= \ent{\rv y} - \frac 1n \condEnt[big]{\ind{\SSS_\AAA}{}}{g_\AAA(\hrv w_\AAA)}  - \frac 1n \condEnt[big]{\rvt y}{g_\AAA(\hrv w_\AAA), \ind{\SSS_\AAA}{}}\\
                                       & \ge \ent{\rv y} - \eps\prm - \frac 1n (1-\eps) \condEnt[big]{\rvt y}{g_\AAA(\hrv w_\AAA), \SSS_\AAA} - \eps \ent{\rv y} \,\,\,\\
                                       &\ge \ent{\rv y} - \eps\prm - \frac 1n \condEnt[big]{\rvt y}{g_\AAA(\hrv w_\AAA), \SSS_\AAA}\\
                                       &\ge \ent{\rv y} - \eps\prm - \frac 1n \sum_{\vt u_\AAA} 
                                         \Pcond[][][big]{g_\AAA(\hrv w_\AAA) = \vt u_\AAA}{\SSS_\AAA} \log\card{\condTyp[][\eps]{\rv y}{\rv u_\AAA}{\vt u_\AAA}} \label{eq:md:typ_bound0} \\
                                       &\ge \ent{\rv y} - \condEnt{\rv y}{\rv u_\AAA} - \eps\prm \label{eq:md:typ_bound} \\
                                       &=  \mutInf{\rv u_\AAA}{\rv y} - \eps\prm . \label{eq:md:mut_bound1}
\end{align}
Here, \cref{eq:data_proc1,eq:data_proc2} follow from the data processing inequality~\cite[Theorem~2.8.1]{Cover2006Elements}, we applied the entropy bound \cite[Theorem~2.6.4]{Cover2006Elements} in \cref{eq:md:typ_bound0}, and the cardinality bound for the set of conditionally typical sequences \cite[Lemma~3.1.(v)]{Han1980Unified} in \cref{eq:md:typ_bound}. In particular, for $\AAA = \JJJ$ we obtain
\begin{align}
  \frac 1n \mutInf{\rvt y}{\rv w_\JJJ} &\ge \mutInf{\rv u_\JJJ}{\rv y} - \eps\prm \stackrel{\mathclap{\cref{eq:md:muJ}}}{\ge} \nu_0 - \eps\prm . \label{eq:md:nu0_achievable}
\end{align}
For $\jzidx \in \JJJ$ and $\AAA = \osup \jzidx$ we obtain the following chain of inequalities, where \cref{eq:md:ratesum,eq:md:typ_bound2} will be justified subsequently.
\begin{align}
  \frac 1n \mutInf{\rvt y}{\rv w_{\jzidx}} &\ge \frac 1n \mutInf[big]{\rvt y}{\hrv w_{\jzidx}} \ge \frac 1n \condMutInf[big]{\rvt y}{\hrv w_{\jzidx}}{\hrv w_\AAA} \\
                                        &= \frac 1n \mutInf[big]{\rvt y}{\hrv w_{\jzidx}\hrv w_\AAA} - \frac 1n \mutInf[big]{\rvt y}{\hrv w_\AAA}\\
                                        & \stackrel{\mathclap{\cref{eq:md:mut_bound1}}}{\ge} \mutInf{\rv u_\AAA\rv u_{\jzidx}}{\rv y} - \eps\prm - \frac 1n \mutInf[big]{\rvt y}{\hrv w_\AAA}\\
                                        &= \mutInf{\rv u_\AAA\rv u_{\jzidx}}{\rv y} - \eps\prm - \frac 1n \mutInf[big]{\rvt x_\AAA}{\hrv w_\AAA} + \frac 1n \condMutInf[big]{\rvt x_\AAA}{\hrv w_\AAA}{\rvt y} \label{eq:md:markov} \\
                                        &\ge  \mutInf{\rv u_\AAA\rv u_{\jzidx}}{\rv y} - \eps\prm - \mutInf{\rv x_\AAA}{\rv u_\AAA} + \condEnt{\rv x_\AAA}{\rv y} - \frac 1n \condEnt[big]{\rvt x_\AAA}{\hrv w_\AAA, \rvt y} \label{eq:md:ratesum} \\
                                        &\ge  \mutInf{\rv u_\AAA\rv u_{\jzidx}}{\rv y} - \eps\prm - \mutInf{\rv x_\AAA}{\rv u_\AAA} +  \condEnt{\rv x_\AAA}{\rv y} - \condEnt{\rv x_\AAA}{\rv u_\AAA, \rv y} \label{eq:md:typ_bound2}\\
                                                                                &= \condMutInf{\rv u_{\jzidx}}{\rv y}{\rv u_\AAA} - \eps\prm \\
                                        &\stackrel{\mathclap{\cref{eq:md:mu}}}{\ge} \nu_{\jzidx} - \eps\prm . \label{eq:md:nuj_achievable}
\end{align}
Equality in \cref{eq:md:markov} follows from the Markov chain $\hrv w_\AAA \mkv \rvt x_\AAA \mkv \rvt y$.
In \cref{eq:md:ratesum}, we used that for $\eps$ small enough and $n$ large enough, we have
\begin{align}
  \frac 1n \mutInf[big]{\rvt x_\AAA}{\hrv w_\AAA} &= \frac 1n \ent[big]{\hrv w_\AAA} \\
                                                  &\le \frac 1n \sum_{j \in \AAA} \ent[big]{\hrv w_j} \label{eq:cond_entr_red} \\
                                                  &\le \sum_{j \in \AAA} \Big( \condMutInf{\rv u_j}{\rv x_j}{\rv u_{\osup j}} + \eps \Big) \label{eq:Rj_def} \\
                                                  &\le \mutInf{\rv u_\AAA}{\rv x_\AAA} + \eps\prm , 
\end{align}
where \cref{eq:cond_entr_red} follows from the chain rule for entropy~\cite[Theorem~2.2.1]{Cover2006Elements} and the data processing inequality~\cite[Theorem~2.8.1]{Cover2006Elements} and \cref{eq:Rj_def} follows from the entropy bound \cite[Theorem~2.6.4]{Cover2006Elements} and the fact that $f_\JJJ$ is an $(n, \hat R_\JJJ + \eps)$-code.
The inequality \cref{eq:md:typ_bound2} can be derived similar to \cref{eq:md:mut_bound1}, as for $n$ large enough and $\eps$ small enough,
\begin{align}
  &\frac 1n \condEnt[big]{\rvt x_\AAA}{\hrv w_\AAA, \rvt y} \le \frac 1n \condEnt[big]{\rvt x_\AAA}{g_\AAA(\hrv w_\AAA), \rvt y}  \\
  &\qquad\qquad \le \frac 1n \condEnt[big]{\rvt x_\AAA, \ind{\SSS_\AAA}{}}{g_\AAA(\hrv w_\AAA), \rvt y} \\
  &\qquad\qquad \le \eps\prm + \frac 1n \condEnt[big]{\rvt x_\AAA}{g_\AAA(\hrv w_\AAA), \rvt y, \SSS_\AAA} \\
  &\qquad\qquad \le \eps\prm + \frac 1n \sum_{\vt u_\AAA, \vt y} 
    \Pcond[][][big]{g_\AAA(\hrv w_\AAA) = \vt u_\AAA, \rvt y = \vt y}{\SSS_\AAA} \log\card{\condTyp[][\eps]{\rv x_\AAA}{\rv u_\AAA, \rv y}{\vt u_\AAA, \vt y}} \\
  &\qquad\qquad \le \eps\prm + \condEnt{\rv x_\AAA}{\rv u_\AAA, \rv y} .
\end{align}
For ${\jzidx} \in \III$, we have similarly to \cref{eq:md:mut_bound1} that
\begin{align}
  \frac 1n \mutInf{\rvt y}{\rv w_{\jzidx}} &\ge \frac 1n \mutInf[big]{\rvt y}{g_{\jzidx}(\rv w_{\jzidx})} \\
                                           &= \ent{\rv y} - \frac 1n \condEnt[big]{\rvt y}{g_{\jzidx}(\rv w_{\jzidx})}\\
                                           &\ge \ent{\rv y} - \frac 1n \condEnt[big]{\rvt y, \ind{\SSS_{\jzidx}}{}}{g_{\jzidx}(\rv w_{\jzidx})}\\
                                           &\ge \ent{\rv y} - \frac 1n \ent[big]{\ind{\SSS_{\jzidx}}{}} - \frac 1n \condEnt[big]{\rvt y}{g_{\jzidx}(\rv w_{\jzidx}), \ind{\SSS_{\jzidx}}{}}\\
                                           &\ge \ent{\rv y} - \eps\prm - \frac 1n \condEnt[big]{\rvt y}{g_{\jzidx}(\rv w_{\jzidx}), \SSS_{\jzidx}}\\
                                           &\ge \ent{\rv y} - \eps\prm - \frac 1n \sum_{\vt u_{\jzidx}} \Pcond{g_\AAA(\rv w_{\jzidx}) = \vt u_{\jzidx}}{\SSS_{\jzidx}} \log\card{\condTyp[n]{\rv y}{\rv u_{\jzidx}}{\vt u_{\jzidx}}} \\
                                           &\ge \ent{\rv y} - \eps\prm - \condEnt{\rv y}{\rv u_{\jzidx}} \\
                                           &= \mutInf{\rv u_{\jzidx}}{\rv y} - \eps\prm \stackrel{\mathclap{\cref{eq:md:mu2}}}{\ge} \nu_{\jzidx} - \eps\prm . \label{eq:md:nuj2_achievable}
\end{align}
Combining \cref{eq:md:nu0_achievable,eq:md:nuj_achievable,eq:md:nuj2_achievable} we see that $(\nu_0 + \eps\prm, \nu_\JJJ + \eps\prm, R_\JJJ + \eps) \in \RRRmi$, completing the proof as $\eps$, $\eps\prm$ were arbitrary.
\end{proof}

\begin{lemma}
  \label{lem:mdceo:converse}
  If $(\nu_0, \nu_\JJJ, R_\JJJ) \in \RRRmi$, then for all $i \in \JJJ$ and $\AAA \subseteq \JJJ$,
  \begin{align}
    \sum_{j \in \AAA} R_j - \nu_i &\ge \condMutInf{\rv x_{\AAA}}{\rv u_{\AAA}}{\rv y \rv q} - \condMutInf{\rv y}{\rv u_{i \setminus \AAA}}{\rv q} ,\\
    \sum_{j \in \AAA} R_j &\ge \condMutInf{\rv x_{\AAA}}{\rv u_{\AAA}}{\rv y \rv q} , \text { and } \\
    \sum_{j \in \AAA} R_j - \nu_0 &\ge \condMutInf{\rv x_{\AAA}}{\rv u_{\AAA}}{\rv y \rv q} - \condMutInf{\rv y}{\rv u_{\JJJ \setminus \AAA}}{\rv q} ,
  \end{align}
  for some random variables $(\rv u_\JJJ, \rv q) \in \PrIn$.
\end{lemma}
\begin{proof}
  For $(\nu_0, \nu_\JJJ, R_{\JJJ}) \in \RRRmi$ we apply \cref{def:ceo_achievable}, choosing an $(n, R_{\JJJ})$-code $f_{\JJJ}$ for $\rv x_{\JJJ}$ and define $\rv u_j \defas f_j(\rvt x_j)$ for $j \in {\JJJ}$. In the following, let either $\AAA = \JJJ$, or $\AAA = \{j\}$ for $j \in \JJJ$. Slightly abusing notation, we define $\nu_\AAA \defas \nu_j$ for $\AAA = \{j\}$ and $\nu_\AAA \defas \nu_0$ for $\AAA=\JJJ$. We thus have
\begin{align}
  \label{eq:proof:mdceo:converse1}
  \frac{1}{n} \mutInf{\rv u_\AAA}{\rvt y} &\ge \nu_\AAA .
\end{align}
With $\rv u_{j,i} \defas (\rv u_j, \rvt x_{j,1}^{i-1})$ and $\rv q_i \defas (\rvt y^{i-1}, \rvt y_{i+1}^n)$ we have for every $\BBB \subseteq \JJJ$,
\begin{align}
  n\sum_{j \in \BBB} R_j &\ge \ent{\rv u_{\BBB}} \\
                         &=\mutInf{\rv u_{\BBB}}{\rvt x_\BBB} \\
                         &=\mutInf{\rv u_{\BBB}}{\rvt x_\BBB, \rvt y} \\
                         &=\mutInf{\rv u_{\BBB}}{\rvt y} + \condMutInf{\rv u_{\BBB}}{\rvt x_\BBB}{\rvt y} \\
                         &=\mutInf{\rv u_{\AAA} \rv u_\BBB}{\rvt y} - \condMutInf{\rv u_{\AAA\setminus\BBB}}{\rvt y}{\rv u_\BBB} + \condMutInf{\rv u_{\BBB}}{\rvt x_\BBB}{\rvt y} \\
                         &=\mutInf{\rv u_{\AAA}}{\rvt y} + \condMutInf{\rv u_{\BBB\setminus\AAA}}{\rvt y}{\rv u_{\AAA}} - \condMutInf{\rv u_{\AAA\setminus\BBB}}{\rvt y}{\rv u_\BBB} + \condMutInf{\rv u_{\BBB}}{\rvt x_\BBB}{\rvt y} \\
                         &\RL[\cref{eq:proof:mdceo:converse1}]\ge n\nu_{\AAA} + \condMutInf{\rv u_{\BBB\setminus\AAA}}{\rvt y}{\rv u_{\AAA}} - \condMutInf{\rv u_{\AAA\setminus\BBB}}{\rvt y}{\rv u_\BBB} + \condMutInf{\rv u_{\BBB}}{\rvt x_\BBB}{\rvt y} \\
                         &\ge n\nu_{\AAA} - \mutInf{\rv u_{\AAA\setminus\BBB}}{\rvt y} + \condMutInf{\rv u_{\BBB}}{\rvt x_\BBB}{\rvt y} \\
                         &= \sum_{i=1}^n \big[ \nu_{\AAA} - \condMutInf{\rv u_{\AAA\setminus\BBB}}{\rv y_i}{\rvt y^{i-1}} + \condMutInf{\rv u_{\BBB}}{\rv x_{\BBB,i}}{\rvt y \rvt x_{\BBB}^{i-1}} \big] \\
                         &\ge \sum_{i=1}^n \big[ \nu_{\AAA} - \condMutInf{\rv u_{\AAA\setminus\BBB, i}}{\rv y_i}{\rv q_i} + \condMutInf{\rv u_{\BBB}}{\rv x_{\BBB,i}}{\rvt y \rvt x_{\BBB}^{i-1}} \big] \\
                         &= \sum_{i=1}^n \big[ \nu_{\AAA} - \condMutInf{\rv u_{\AAA\setminus\BBB, i}}{\rv y_i}{\rv q_i} + \condMutInf{\rv u_{\BBB, i}}{\rv x_{\BBB,i}}{\rv y_i \rv q_i} \big] .
\end{align}
The result now follows by a standard time-sharing argument. Note that the required Markov chains and the independence constraints are satisfied.
\end{proof}

The following result is a simple corollary of \cref{lem:mdceo:converse} and will suffice for us.
\begin{corollary}
  \label{cor:mdceo:converse}
  For any $(\nu_0, \nu_\JJJ, R_\JJJ) \in \RRRmi$ there are random variables $(\rv u_\JJJ, \rv q) \in \PrIn$ with
  \begin{align}
    R_j &\ge 0, &&\shiftleftB\text{for all } j \in \JJJ , \label{eq:md:ro1} \\
    \sum_{j \in \AAA} R_j - \nu_0 &\ge \condMutInf{\rv x_{\AAA}}{\rv u_{\AAA}}{\rv y \rv q} - \condMutInf{\rv y}{\rv u_{\JJJ \setminus \AAA}}{\rv q}, &&\shiftleftB\text{for all } \AAA \subseteq \JJJ , \label{eq:md:ro2} \\
    R_j - \nu_j &\ge \condMutInf{\rv x_j}{\rv u_j}{\rv y \rv q}, &&\shiftleftB\text{for all } j \in \JJJ , \label{eq:md:ro3} \\
    \nu_j &\le \condMutInf{\rv y}{\rv u_j}{\rv q}, &&\shiftleftB\text{for all } j \in \JJJ . \label{eq:md:ro4}
  \end{align}
\end{corollary}

In the following proof, we will make use of some rather technical results on convex polyhedra, derived in \cref{apx:results-conv-polyh}.
\begin{proof*}[Proof of \cref{thm:mdceo:main}]
  Assume $(\nu_0, \nu_\JJJ, R_\JJJ) \in \RRRmi$. We can then find $(\rv u_\JJJ, \rv q) \in \PrIn$ such that \cref{eq:md:ro1,eq:md:ro2,eq:md:ro3,eq:md:ro4} hold.
  We define $(\wt\nu_0, \wt\nu_\JJJ) \defas - (\nu_0, \nu_\JJJ)$ to simplify notation. It is straightforward to check that the inequalities \cref{eq:md:ro1,eq:md:ro2,eq:md:ro3,eq:md:ro4} define a sequence of closed convex polyhedra $\HHH^{(j)}$ in the variables $(R_\JJJ, \wt{\nu}_0, \wt\nu_\JJJ)$ that satisfy \cref{itm:recursion,itm:line-free} of \cref{lem:convex_polyhedra}. $\HHH^{(0)}$ is defined by \cref{eq:md:ro1,eq:md:ro2} alone, and for $\jzidx \in \NZtoo{J}$ the polyhedron $\HHH^{(\jzidx)}$ is given in the $K+\jzidx$ variables $(R_\JJJ, \wt{\nu}_0, \wt{\nu}_1, \wt{\nu}_2, \dots, \wt{\nu}_{\jzidx})$ by adding constraints \cref{eq:md:ro3,eq:md:ro4} for each $j \in \Ntoo{\jzidx}$.
  The set $\HHH^{(0)}$ is a supermodular polyhedron \cite[Section~2.3]{Fujishige2005Submodular} in the $K$ variables $(R_\JJJ, \wt{\nu}_0)$ on $(\KKK, 2^{\KKK})$ with rank function
  \begin{align}
    \vartheta(\AAA) =
    \begin{cases}
      0, & K \notin \AAA, \\
      \condMutInf{\rv x_{\AAA\setminus K}}{\rv u_{\AAA \setminus K}}{\rv y \rv q} - \condMutInf{\rv y}{\rv u_{\JJJ \setminus \AAA}}{\rv q}, & K \in \AAA,
    \end{cases} 
  \end{align}
  where supermodularity follows via standard information-theoretic arguments.
  By the extreme point theorem \cite[Theorem~3.22]{Fujishige2005Submodular}, every extreme point of $\HHH^{(0)}$ is associated with a total order $\osub$ on $\KKK$. Such an extreme point is given by
  \begin{align}
    R_j^{(\osub)} &= 0 \text{ for } j \sqsubset K , \eqnl
    R_j^{(\osub)} &= \condMutInf{\rv u_j}{\rv x_j}{\rv u_{\osup j} \rv q} \text{ for } j \sqsupset K , \eqnl
    \nu_0^{(\osub)} &= \condMutInf{\rv y}{\rv u_{\osup K}}{\rv q} - \condMutInf{\rv y}{\rv u_{\osub K}}{\rv y \rv q} .
  \end{align}
  \Cref{itm:one-cut} of \cref{lem:convex_polyhedra} is now verified by
  \begin{align}
    R_j^{(\osub)} \le \condMutInf{\rv x_j}{\rv u_j}{\rv y \rv q} + \condMutInf{\rv y}{\rv u_j}{\rv q} = \condMutInf{\rv x_j}{\rv u_j}{\rv q} .
  \end{align}
  By applying \cref{lem:convex_polyhedra} we find that every extreme point of $\HHH^{(J)}$ is given by a subset $\III \subseteq \JJJ$ and an order $\osub$ of $\KKK$ as
  \begin{align}
    R_j^{(\osub, \III)} &= \condMutInf{\rv x_{j}}{\rv u_{j}}{\rv q}, &&\shiftleftC j \in \III, \label{eq:ext_pt1} \\
    R_j^{(\osub, \III)} &= 0, &&\shiftleftC j \notin \III \text{ and } j \sqsubset K, \label{eq:ext_pt2} \\
    R_j^{(\osub, \III)} &= \condMutInf{\rv u_j}{\rv x_j}{\rv u_{\osup j} \rv q}, &&\shiftleftC j \notin \III \text{ and } j \sqsupset K, \label{eq:ext_pt3} \\
    \nu_j^{(\osub, \III)} &= \condMutInf{\rv u_{j}}{\rv y}{\rv q} ,&&\shiftleftC j \in \III, \label{eq:ext_pt5} \\
    \nu_j^{(\osub, \III)} &= - \condMutInf{\rv u_{j}}{\rv x_{j}}{\rv y \rv q} ,&&\shiftleftC j \notin \III \text{ and } j \sqsubset K, \label{eq:ext_pt6} \\
    \nu_j^{(\osub, \III)} &= \condMutInf{\rv u_{j}}{\rv y}{\rv u_{\osup {j}} \rv q} ,&&\shiftleftC j \notin \III \text{ and } j \sqsupset K , \label{eq:ext_pt7} \\
    \nu_0^{(\osub, \III)} &= \condMutInf{\rv y}{\rv u_{\osup K}}{\rv q} - \mathrlap{\condMutInf{\rv y}{\rv u_{\osub K}}{\rv y \rv q} .} \label{eq:ext_pt4}
  \end{align}
  For each $q \in \QQQ$ with $\Prob{\rv q = q} > 0$ let the point $(\nu_0^{(\osub, \III, q)}, \nu_\JJJ^{(\osub, \III, q)}, R_\JJJ^{(\osub, \III, q)})$ be defined by \cref{eq:ext_pt1,eq:ext_pt2,eq:ext_pt3,eq:ext_pt4,eq:ext_pt5,eq:ext_pt6,eq:ext_pt7}, but given $\{\rv q = q\}$. By substituting $\rv u_j \rightarrow \varnothing$ if $j \notin \III \text{ and } j \sqsubset K$, we see that $(\nu_0^{(\osub, \III, q)}, \nu_\JJJ^{(\osub, \III, q)}, R_\JJJ^{(\osub, \III, q)}) \in \RRRmi^{(\osub, \III)}$ and consequently $(\nu_0^{(\osub, \III)}, \nu_\JJJ^{(\osub, \III)}, R_\JJJ^{(\osub, \III)}) \in \cvx{\RRRmi^{(\osub, \III)}}$.
  Defining the orthant $\OOO \defas \{ (\nu_0, \nu_\JJJ, R_\JJJ) : \nu_0 \le 0, \nu_\JJJ \le \vt 0, R_\JJJ \ge \vt 0 \}$, this implies
  \begin{align}
    (\nu_\KKK, R_\JJJ) \in \cvx{\bigcup_{\osub, \III} \cvx{\RRRmi^{(\osub, \III)}}} + \OOO &= \cvx{\bigcup_{\osub, \III} \RRRmi^{(\osub, \III)}} + \cvx{\OOO}\\
                                                                                                        &= \cvx{\bigcup_{\osub, \III} \RRRmi^{(\osub, \III)} + \OOO} \label{eq:ext_pt_sum} \\
                                                                                                        &= \cvx{\bigcup_{\osub, \III} \RRRmi^{(\osub, \III)}} \label{eq:ext_pt_plusorth} ,
  \end{align}
  where \cref{eq:ext_pt_sum} follows from \cite[Theorem~1.1.2]{Schneider2014Convex} and in \cref{eq:ext_pt_plusorth} we used that $\RRRmi^{(\osub, \III)} + \OOO = \RRRmi^{(\osub, \III)}$ by definition.
  Together with \cref{lemma:mdceo:achiev} and the convexity of $\ol{\RRRmi}$ we obtain
  \begin{align}
    \RRRmi \subseteq \cvx{ \bigcup_{\osub, \III} \RRRmi^{(\osub, \III)} } \subseteq \ol{\RRRmi}.    
  \end{align}
  Note that $\ol{\RRRmi}$ is convex by a time-sharing argument.

  It remains to show that $\cvx{\bigcup_{\osub, \III} \RRRmi^{(\osub, \III)}}$ is closed. Using \cref{pro:mdceo:cardinality}, we can write $\RRRmi^{(\osub, \III)} = \vt F^{(\osub, \III)}(\PrIn\prm) + \OOO$, where $\PrIn\prm \defas \{\p[\rv y, \rv x_\JJJ, \rv u_\JJJ]{} : (\rv u_\JJJ, \varnothing) \in \PrIn, \card{\UUU_j} = \card{\XXX_j} + 4^J, j \in \JJJ\}$ is a compact subset of the probability simplex and $\vt F^{(\osub, \III)}$ is a continuous function, given by the definition of $\RRRmi^{(\osub, \III)}$, \cref{eq:md:rate,eq:md:rate2,eq:md:mu,eq:md:mu2,eq:md:muJ}. We can thus write
  \begin{align}
    \cvx{\bigcup_{\osub, \III} \RRRmi^{(\osub, \III)}} &= \cvx{\bigcup_{\osub, \III} \vt F^{(\osub, \III)}(\PrIn\prm) + \OOO } ,
  \end{align}
  which is closed by \cref{lem:convex_sum}.
\end{proof*}

\subsection{Proof of \texorpdfstring{\cref{pro:mdceo:cardinality}}{Proposition \ref{pro:mdceo:cardinality}}}
\label{sec:proof:mdceo:cardinality}

Pick arbitrary $j,\jzidx \in \JJJ$. For nonempty $\BBB \subseteq \JJJ$ with $j \in \BBB$ we can write $\condEnt{\rv x_{\jzidx}}{\rv u_\BBB} = \Exp[\rv u_j][big]{f_{\jzidx,\BBB}\big(\pcond[\rv x_j][\rv u_j][normal]{\wc}{\rv u_j}\big)}$ as well as $\condEnt{\rv y}{\rv u_\BBB} = \Exp[\rv u_j][big]{g_{\BBB}\big(\pcond[\rv x_j][\rv u_j][normal]{\wc}{\rv u_j}\big)}$, where 
\begin{align}
  f_{\jzidx,\BBB}\big(\pcond[\rv x_j][\rv u_j][normal]{\wc}{u_j}\big) &\defas \condEnt{\rv x_{\jzidx}}{\rv u_{\BBB \setminus j}, \rv u_j = u_j} , \\
  g_{\BBB}\big(\pcond[\rv x_j][\rv u_j][normal]{\wc}{u_j}\big) &\defas \condEnt{\rv y}{\rv u_{\BBB \setminus j}, \rv u_j = u_j} .  
\end{align}
Observe that $f_{\jzidx,\BBB}$ and $g_{\BBB}$ are continuous functions of $\pcond[\rv x_j][\rv u_j]{\wc}{u_j}$.
Apply the support lemma \cite[Appendix~C]{ElGamal2011Network} with the functions $f_{\jzidx,\BBB}$ and $g_{\BBB}$ for all $\jzidx \in \JJJ$, $j \in \BBB \subseteq \JJJ$, and $\card{\XXX_j} - 1$ test functions, which guarantee that the marginal distribution $\p[\rv x_j]{}$ does not change. We obtain a new random variable $\hrv u_j$ with $\condEnt[normal]{\rv x_{\jzidx}}{\rv u_{\BBB\setminus j} \hrv u_j} = \condEnt[normal]{\rv x_{\jzidx}}{\rv u_\BBB}$ and $\condEnt[normal]{\rv y}{\rv u_{\BBB\setminus j} \hrv u_j} = \condEnt[normal]{\rv y}{\rv u_{\BBB}}$. By rewriting~\cref{eq:md:rate,eq:md:rate2,eq:md:mu,eq:md:mu2,eq:md:muJ} in terms of conditional entropies, it is evident that the defining inequalities for $\RRRmi^{(\osub, \III)}$ remain the same when replacing $\rv u_j$ by $\hrv u_j$.
The support of $\hrv u_j$ satisfies the required cardinality bound\footnote{There are $J$ ways to choose $\jzidx$ and $2^{J-1}$ ways to choose $\BBB$.}
\begin{align}
  \card[normal]{\hat\UUU_j} &\le \card{\XXX_j} - 1 + J 2^{J-1} + 2^{J-1} \\
                    &\le \card{\XXX_j} + 4^J .
\end{align}
The same process is repeated for every $j \in \JJJ$.

\subsection{Results on convex polyhedra}
\label{apx:results-conv-polyh}

We start this \lcnamecref{apx:results-conv-polyh} with a simple \lcnamecref{lem:convex_sum}, which will be used in several proofs.

\begin{lemma}
  \label{lem:convex_sum}
  For a compact set $\CCC \subseteq \RR^n$ and a closed, convex set $\BBB \subseteq \RR^n$,
  \begin{align}
    \AAA \defas \cvx{\CCC + \BBB} = \cvx{\CCC} + \BBB = \ol{\AAA} .
  \end{align}
\end{lemma}
\begin{proof}
  We have $\AAA = \cvx{\CCC} + \cvx{\BBB} = \cvx{\CCC} + \BBB$ by \cite[Theorem~1.1.2]{Schneider2014Convex} and the convexity of $\BBB$. Note that $\cvx{\CCC}$ is compact by \cite[Theorem~2.3.4]{Gruenbaum2003Convex}. $\AAA$ is the sum of a compact set and a closed set and, hence, closed \cite[Exercise~1.3(e)]{Rudin1991Functional}.
\end{proof}

Let $\HHH$ be the convex polyhedron $\HHH \defas \{ \vt x \in \RR^n: \vt A \vt x \ge \vt b \}$ for an $m \times n$ matrix $\vt A = (\vt a_{(1)}, \vt a_{(2)}, \dots, \vt a_{(m)})\transp$ and $\vt b \in \RR^m$, where $\vt a_{(j)}\transp$ is the $j$th row of $\vt A$.
In this section we will use the notation of \cite{Gruenbaum2003Convex}. In particular, we shall call a closed convex set \emph{line-free} if it does not contain a (straight) line. The \emph{characteristic cone} of a closed convex set $\CCC$ is defined as $\cc{\CCC} \defas \{\vt y : \vt x + \lambda\vt y \in \CCC \text{ for all } \lambda \ge 0\}$ ($\vt x \in \CCC$ arbitrary) and $\ext{\CCC}$ is the set of all \emph{extreme points} of $\CCC$, \ie, points $\vt x \in \CCC$ that cannot be written as $\vt x = \lambda \vt y + (1-\lambda) \vt z$ with $\vt y, \vt z \in \CCC$, $\vt y \neq \vt z$ and $\lambda \in (0,1)$.

\begin{lemma}
  \label{lem:char_cone1}
  A point $\vt y$ is in $\cc{\HHH}$ if and only if $\vt A \vt y \ge \vt 0$.
\end{lemma}
\begin{proof}
  If $\vt A \vt y \ge \vt 0$, $\vt x \in \HHH$ and $\lambda \ge 0$, $\vt A(\vt x + \lambda \vt y) \ge \vt A \vt x \ge \vt b$. On the other hand, for $\vt a_{(j)}\transp \vt y < 0$, we have $\vt a_{(j)}\transp (\vt x + \lambda \vt y) < b_j$ for $\lambda > \frac{b_j - \vt a_{(j)}\transp \vt x}{\vt a_{(j)}\transp \vt y} > 0$.
\end{proof}

\begin{lemma}
  \label{lem:char_cone2}
   If, for every $i \in \Ntoo{n}$, there exists $j \in \Ntoo{m}$ such that $\vt e_i = \vt a_{(j)}$ and for every $j \in \Ntoo{m}$, $\vt a_{(j)} \ge \vt 0$, then $\HHH$ is line-free and $\cc{\HHH} = \RR_+^n$.
\end{lemma}
\begin{proof}
  For any $\vt y \in \RR_+^n$, clearly $\vt A\vt y \ge \vt 0$ and hence $\vt y \in \cc{\HHH}$ by \cref{lem:char_cone1}. If $\vt y \notin \RR_+^n$ we have $y_i < 0$ for some $i \in \Ntoo{n}$ and choose $j \in \Ntoo{m}$ such that $\vt a_{(j)} = \vt e_i$, resulting in $\vt a_{(j)}\transp \vt y = y_i < 0$.
  To show that $\HHH$ is line-free assume that $\vt x + \lambda \vt y \in \HHH$ for all $\lambda \in \RR$. This implies $\pm \vt y \in \cc{\HHH}$, \ie, $\vt y = \vt 0$.
\end{proof}

\begin{definition}
  A point $\vt x$ is on an \emph{extreme ray} of the cone $\cc{\HHH}$ if the decomposition $\vt x = \vt y + \vt z$ with $\vt y, \vt z \in \cc{\HHH}$ implies that $\vt y=\lambda \vt z$ for some $\lambda \in \RR$.
\end{definition}

It is easy to see that the points on extreme rays of $\OOO$ are given by $\vt x = \lambda \vt e_i$ for $\lambda \ge 0$ and $i \in \Ntoo{n}$.

Define $\AAA(\vt x) \defas \{ j \in \Ntoo{m} : \vt a_{(j)}\transp \vt x = b_j \}$. We say that exactly $n_0$ linearly independent inequalities are satisfied with equality at $\vt x$, if $\vt A \vt x \ge \vt b$ and $(\vt a_{(j)})_{j \in \AAA(\vt x)}$ has rank $n_0$.

\begin{lemma}
  \label{lem:extreme_point_inequalities}
  $\vt x \in \ext{\HHH}$ if and only if exactly $n$ linearly independent inequalities are satisfied with equality at $\vt x$.
\end{lemma}
\begin{proof}
  Assuming that less than $n$ linearly independent inequalities are satisfied with equality at $\vt x$, we find $\vt 0 \neq \vt c \perp (\vt a_{(j)})_{j \in \AAA(\vt x)}$ and thus  $\vt x \pm \eps\vt c \in \HHH$ for a small $\eps > 0$, showing that $\vt x \notin \ext{\HHH}$.

  Conversely assume $\vt x \notin \ext{\HHH}$, \ie, $\vt x = \lambda \vt x\prm + (1-\lambda)\vt x\pprm$ for $\lambda \in (0,1)$ and $\vt x\prm, \vt x\pprm \in \HHH$, $\vt x\prm \neq \vt x\pprm$. For any $j \in \AAA(\vt x)$, we then have $\lambda \vt a_{(j)}\transp \vt x\prm + (1-\lambda) \vt a_{(j)}\transp \vt x\pprm = b_j$, which implies $\vt a_{(j)}\transp \vt x\prm = \vt a_{(j)}\transp \vt x\pprm = b_j$ and therefore $\vt 0 \neq \vt x\prm - \vt x\pprm \perp (\vt a_{(j)})_{j \in \AAA(\vt x)}$.
\end{proof}

\begin{lemma}
  \label{lem:extreme_line_inequalities}
  Assuming that $\HHH$ is line free and that exactly $n-1$ linearly independent inequalities are satisfied with equality at $\vt x$. Then either $\vt x = \lambda \vt c + (1-\lambda) \vt d$ where $\lambda \in (0,1)$ and $\vt c, \vt d \in \ext{\HHH}$ or $\vt x = \vt c + \vt d$ where $\vt c \in \ext{\HHH}$ and $\vt d \neq \vt 0$ lies on an extreme ray of $\cc{\HHH}$.
\end{lemma}
\begin{proof}
  We obtain $\vt 0 \neq \vt r \perp (\vt a_{(j)})_{j \in \AAA(\vt x)}$. Define $\lambda_1 \defas \inf \{\lambda : \vt x + \lambda \vt r \in \HHH \}$ and $\lambda_2 \defas \sup \{\lambda : \vt x + \lambda \vt r \in \HHH \}$. Clearly $\lambda_1 \le 0 \le \lambda_2$. As $\HHH$ is line free, we may assume without loss of generality $\lambda_1 = -1$ (note that $\vt x \notin \ext{\HHH}$) and set $\vt c = \vt x - \vt r$. We now have $\vt c \in \ext{\HHH}$ as otherwise $\vt c - \eps \vt r \in \HHH$ for some small $\eps > 0$.

  If $\lambda_2 < \infty$, define $\vt d = \vt x + \lambda_2 \vt r$, which yields $\vt d \in \ext{\HHH}$ and $\vt x = \lambda \vt c + (1-\lambda) \vt d$ with $\lambda = \frac{\lambda_2}{\lambda_2 + 1}$. Note that $\lambda_2 \neq 0$ as $\vt x \notin \ext{\HHH}$.

  If $\lambda_2 = \infty$ we have $\vt x-\vt c = \vt r \in \cc{\HHH}$. We need to show that $\vt r$ is also on an extreme ray of $\cc{\HHH}$. Assuming $\vt r = \vt r\prm + \vt r\pprm$ with $\vt r\prm, \vt r\pprm \in \cc{\HHH}$ yields $\vt a_{(j)}\transp (\vt r\prm + \vt r\pprm) = 0$, which implies $\vt a_{(j)}\transp \vt r\prm = \vt a_{(j)}\transp \vt r\pprm = 0$ for every $j \in \AAA(\vt x)$ by \cref{lem:char_cone1}.
\end{proof}

For each $j \in \NZtoo{J}$, define the closed convex polyhedron $\HHH^{(j)} \defas \{ \vt x \in \RR^{K+j} : 
\vt A^{(j)} \vt x \ge \vt b^{(j)} \}$, where $\vt A^{(j)}$ is a matrix and $\vt b^{(j)}$ a vector of appropriate dimension.
We make the following three assumptions:
\begin{enumerate}
\item $\vt A^{(j)}$ and $\vt b^{(j)}$ are defined recursively as
  \begin{align}
    \vt A^{(j)} &\defas
              \begin{pmatrix}
                \vt A^{(j-1)} & \vt 0 \\
                \vt 0\transp  & 1 \\
                \vt e_j\transp  & 1
              \end{pmatrix}, &
                              \vt b^{(j)} &= \begin{pmatrix}
                                \vt b^{(j-1)} \\
                                c^{(j)}_1  \\
                                c^{(j)}_2
                              \end{pmatrix},
  \end{align}
  where $c^{(j)}_1$ and $c^{(j)}_2$ are arbitrary reals.\label[assumption]{itm:recursion}
\item Each entry of $\vt A^{(0)}$ equals $0$ or $1$ and for all $k \in \KKK$ at least one row of $\vt A^{(0)}$ is equal to $\vt e_k\transp$.
  Due to \cref{itm:recursion}, this also implies that each entry of $\vt A^{(j)}$ is in $\{0,1\}$ and for all $k \in \Ntoo{K+j}$ at least one row of $\vt A^{(j)}$ is equal to $\vt e_k\transp$.
  \label[assumption]{itm:line-free}
\item For any extreme point $\vt x \in \ext{\HHH^{(0)}}$ and any $j \in \JJJ$, assume $x_j \le c^{(j)}_2-c^{(j)}_1$.
  \label[assumption]{itm:one-cut}
\end{enumerate}

\begin{lemma}
  \label{lem:convex_polyhedra}
    Under \cref{itm:recursion,itm:line-free,itm:one-cut}, for every $\jzidx \in \NZtoo{J}$ and every extreme point $\vt y \in \ext{\HHH^{(\jzidx)}}$ there is an extreme point $\vt x \in \ext{\HHH^{(0)}}$ and a subset $\III_{\jzidx} \subseteq \Ntoo{\jzidx}$ such that $y_K = x_K$ and for every $j \in \JJJ$,
  \begin{align}
    \label{eq:convex:y1}
    y_j &= 
    \begin{cases}
      x_j, & j \notin \III_{\jzidx}, \\
      c_2^{(j)} - c_1^{(j)}, & j \in \III_{\jzidx},
    \end{cases}
  \end{align}
  and for every $j \in \Ntoo{\jzidx}$,
  \begin{align}
    \label{eq:convex:y2}
    y_{K+j} &=
    \begin{cases}
      c_2^{(j)} - x_j, & j \notin \III_{\jzidx}, \\
      c_1^{(j)}, & j \in \III_{\jzidx}.
    \end{cases}
  \end{align}
\end{lemma}
\begin{proof}
  For every $j \in \JJJ$, $\HHH^{(j)}$ is line free by \cref{itm:line-free} and \cref{lem:char_cone2} and can be written \cite[Lemma~6,~p.~25]{Gruenbaum2003Convex} as $\HHH^{(j)} = \cc{\HHH^{(j)}} + \cvx{\ext{\HHH^{(j)}}}$. \Cref{lem:char_cone2} also implies $\cc{\HHH} ^{(j)} = \OOO$.

  Let us proceed inductively over $\jzidx \in \NZtoo{J}$. For $\jzidx=0$ the statement is trivial.
  Given any $\vt y \in \ext{\HHH^{(\jzidx)}}$, we need to obtain $\vt x \in \ext{\HHH^{(0)}}$ and $\III_{\jzidx}$ such that $\vt y$ is given according to \cref{eq:convex:y1,eq:convex:y2}. Let $\vt z = \vt y_1^{K+\jzidx-1}$ be the truncation of $\vt y$. Exactly $K+\jzidx$ linear independent inequalities of $\vt A^{(\jzidx)}  \vt y \ge \vt b^{(\jzidx)}$ are satisfied with equality by \cref{lem:extreme_point_inequalities}, which is possible in only two different ways:
  \begin{itemize}
  \item \emph{Construction I:} Exactly $K+\jzidx-1$ linear independent inequalities of $\vt A^{(\jzidx-1)}  \vt z \ge \vt b^{(\jzidx-1)}$ are satisfied with equality, \ie, $\vt z \in \ext{\HHH ^{(\jzidx-1)}}$ by \cref{lem:extreme_point_inequalities}, and at least one of 
    \begin{align}
      y_{K+\jzidx} &\ge c^{(\jzidx)}_1,
            \label{eq:convex:cond1} \\*
      y_{\jzidx} + y_{K+\jzidx} &\ge c^{(\jzidx)}_2, \label{eq:convex:cond2} 
    \end{align}
    is satisfied with equality.

    As $\vt z \in \ext{\HHH^{(\jzidx-1)}}$, there exists $\vt x\in \ext{\HHH^{(0)}}$ and $\III_{\jzidx-1}$ such that \cref{eq:convex:y1} holds for $j \in \JJJ$ and \cref{eq:convex:y2} holds for $j \in \Ntoo{\jzidx-1}$ by the induction hypothesis. In particular $y_{\jzidx} = x_{\jzidx}$. Assuming that \cref{eq:convex:cond2} holds with equality, we have $y_{K+\jzidx} = c^{(\jzidx)}_2 - x_{\jzidx}$. Thus, the point $\vt x$ together with $\III_{\jzidx} = \III_{\jzidx-1}$ yields $\vt y$ from \cref{eq:convex:y1,eq:convex:y2}.
    Equality in \cref{eq:convex:cond1} implies equality in \cref{eq:convex:cond2} by \cref{itm:one-cut}.
  \item \emph{Construction II:} Exactly $K+\jzidx-2$ linear independent inequalities of $\vt A^{(\jzidx-1)}  \vt z \ge \vt b^{(\jzidx-1)}$ are satisfied with equality and  \cref{eq:convex:cond1,eq:convex:cond2} are both satisfied with equality as well. Additionally, these $K+\jzidx$ inequalities together need to be linearly independent.
    This can occur in two different ways by \cref{lem:extreme_line_inequalities}.
    
    Assume $\vt z = \lambda \vt x + (1-\lambda) \vt x\prm$ for $\vt x, \vt x\prm \in  \ext{\HHH^{(\jzidx-1)}}$, $\vt x \neq \vt x\prm$ and $\lambda \in (0,1)$. This implies $y_{K+\jzidx} = c^{(\jzidx)}_1$ and $y_{\jzidx} = \lambda x_{\jzidx} + (1-\lambda) x\prm_{\jzidx} = c^{(\jzidx)}_2 - c^{(\jzidx)}_1$, which by \cref{itm:one-cut} already implies $x_{\jzidx} = x\prm_{\jzidx} = c^{(\jzidx)}_2 - c^{(\jzidx)}_1$. Thus, \cref{eq:convex:cond1,eq:convex:cond2} are satisfied (with equality) for every $\lambda \in [0,1]$ and $\vt y$ cannot be an extreme point as it can be written as a non-trivial convex combination.

    We can thus focus on the second option which is that $\vt z$ is on an extreme ray of $\HHH^{(\jzidx-1)}$, \ie, $\vt z = \vt x + \lambda \vt e_{\jzidx\prm}$ for some $\vt x \in \ext{\HHH^{(\jzidx-1)}}$, $\lambda > 0$ and $\jzidx\prm \in \Ntoo{K+\jzidx-1}$. If $\jzidx\prm \neq \jzidx$, \cref{eq:convex:cond1,eq:convex:cond2} are satisfied for all $\lambda > 0$ and thus $\vt y$ cannot be an extreme point because it can be written as a non-trivial convex combination.
    For $\jzidx\prm = \jzidx$ the point $\vt x$ with $\III_{\jzidx} = \III_{\jzidx-1} \cup \jzidx$ yields the desired extreme point. \qedhere
  \end{itemize}
\end{proof}

\vspace*{2ex}\noindent

\bibliographystyle{imaiai}
\bibliography{IEEEabrv,literature}

\end{document}